\documentclass[useAMS,usenatbib,usegraphicx]{mn2e}
\usepackage{amsmath}
\usepackage{color}
\usepackage{subfig}

\newif\ifAMStwofonts
\AMStwofontstrue

\newcommand{\simlt}{\lower.5ex\hbox{$\; \buildrel < \over \sim \;$}}
\newcommand{\simgt}{\lower.5ex\hbox{$\; \buildrel > \over \sim \;$}}

\title[{\sc SHARDS}: Dust attenuation in z$\sim$2 galaxies]
{{\sc SHARDS}: Constraints on the dust attenuation law of star-forming galaxies at z$\sim$2}
\author[M. Tress et al.]
{M\'onica Tress$^1$, 
Esther M\'armol-Queralt\'o$^2$,
Ignacio Ferreras$^1$\thanks{E-mail: i.ferreras@ucl.ac.uk}, \and
Pablo~G. P\'erez-Gonz\'alez$^3$,
Guillermo Barro$^4$,
Bel\'en Alcalde Pampliega$^3$, \and
Antonio Cava$^5$,
Helena Dom\'\i nguez-S\'anchez$^3$,
Carmen Eliche-Moral$^3$, \and
N\'estor Espino-Briones$^3$, 
Pilar Esquej$^6$,
Antonio Hern\'an-Caballero$^3$,\and
Giulia Rodighiero$^7$,
Luc\'\i a Rodriguez-Mu\~{n}oz$^7$\\
$^1$ Mullard Space Science Laboratory, University College London, 
Holmbury St Mary, Dorking, Surrey RH5 6NT, UK\\
$^2$ SUPA, Institute for Astronomy, University of Edinburgh, Royal Observatory, Edinburgh EH9 3HJ, UK\\
$^3$ Departamento de Astrof\'\i sica, Facultad de CC. F\'\i sicas, Universidad Complutense de Madrid, E-28040 Madrid, Spain\\
$^4$ Department of Physics, University of the Pacific, Stockton, CA 95211, USA\\
$^5$ Observatoire de Gen\`eve, Universit\'e de Gen\`eve, 51 Ch. des Maillettes, 1290, Versoix, Switzerland\\
$^6$ Herschel Science Centre, ESA, Villafranca del Castillo, Apartado 78, E-28691 Villanueva de la Ca\~{n}ada, Spain\\
$^7$ Dipartimento di Fisica e Astronomia, Universit\`a di Padova, vicolo dell'Osservatorio 2, I-35122 Padova, Italy
}

\begin{document}
\date{Accepted 2017 December 18. Received 2017 November 29; in original form 2017 September 4.}
\pagerange{\pageref{firstpage}--\pageref{lastpage}} \pubyear{2018}
\maketitle
\label{firstpage}

%%\newif\ifAMStwofonts
%%\AMStwofontstrue

\begin{abstract}
 We make use of {\sc SHARDS}, an
ultra-deep ($<$26.5AB) galaxy survey that provides optical
photo-spectra at resolution R$\sim$50, via medium band filters
(FWHM$\sim$150\AA). This dataset is combined with ancillary optical and
NIR fluxes to constrain the dust attenuation law in the
rest-frame NUV region of star-forming galaxies within the redshift
window 1.5$<$z$<$3. We focus on the NUV bump strength (B) and the
total-to-selective extinction ratio (R$_V$), targeting a sample of
1,753 galaxies. By comparing the data with a set of population
synthesis models coupled to a parametric dust attenuation law, we
constrain R$_V$ and B, as well as the colour excess,
E($B-V$). We find a correlation between R$_V$ and B, that can be
interpreted either as a result of the grain size distribution, or a
variation of the dust geometry among galaxies. According to the
former, small dust grains are associated with a stronger NUV bump. The
latter would lead to a range of clumpiness in the distribution of dust
within the interstellar medium of star-forming galaxies.  The observed
wide range of NUV bump strengths can lead to a systematic in the
interpretation of the UV slope $\beta$ typically used to characterize
the dust content. In this study we quantify these
variations, concluding that the effects are $\Delta\beta\sim$0.4.
\end{abstract} 

\begin{keywords}
galaxies: ISM -- ISM: dust, extinction -- galaxies: stellar content -- galaxies: high-redshift
\end{keywords}

%%%%%%%%%%%%%%%%%%%%%%%%%%%%%%%%%%%%%%%%%%%%%%%%
\section{Introduction}
\label{Sec:Intro}

Interstellar dust is an ubiquitous component in galaxies, affecting
the interpretation of photometric and spectroscopic observations
\citep[see, e.g.][]{Galliano:17}. Made
up of particles with a wide range of sizes, from 0.01\,$\mu$m to
0.2\,$\mu$m \citep[e.g.][]{Draine:03}, it affects light through
scattering and absorption, in a wavelength-dependent manner, with
increased cross sections at shorter wavelength. Consequently,
constraining the dust parameters is important to decipher the
properties of the illuminating source, i.e. the stellar
component. Specifically, dust corrections help derive robust star
formation rates, stellar masses, and stellar population
parameters. Unfortunately, the degeneracy between dust and age
complicates the derivation of robust population parameters.
If the composition and distribution of dust grain sizes is known,
it is possible to calculate the fraction of light lost along the line
of sight. This so-called extinction curve has a fairly featureless
distribution, and can be modelled by a set of power laws.  In the
magnitude scale, one defines the extinction in a given band, $X$ as
the magnitude increase (i.e. dimming) caused by dust in that band:
A$_X$.  A parameter commonly used to characterize the dust attenuation
law is the total-to-selective ratio R$_{V}$=A$_{V}$/E($B-V$), where
A$_V$ is the extinction through the $V$ band and E($B-V$)=A$_B$-A$_V$
is the reddening, or colour excess. A high value of R$_V$ represents
a \textit{grey} law, where A$_B\rightarrow$A$_V$, i.e. there is no
strong wavelength dependence of the extincion law. At the other end, a
low value of R$_V$ represents a very strongly wavelength-dependent
law.

In addition to this general trend, there is a set of
features related to specific components of the dust. In the
NUV/optical region, the most prominent one is the 2,175\AA\ NUV
bump \citep{STP:69}. Possible candidates to explain this feature are
graphite or PAH molecules because this wavelength corresponds to
$\pi\rightarrow\pi^*$ electronic excitations in $sp^2$-bonded carbon
structures \citep{Draine:89} although other candidates are
possible \citep{Bradley:05}. Locally, in the Milky Way and the Large
Magellanic Cloud, the NUV bump is present along different lines of
sight \citep{F:99,BHT:15}. Dust extinction in the Small
Magellanic Cloud has been a prototypical example lacking the
bump \citep{Pei:92}. However, evidence has been recently presented to the
contrary \citep{HSH:16}. In galaxies, the interpretation of the effect
of dust gets more complicated as different stellar populations are
affected in different ways, so that not only dust composition but its
distribution within the galaxy will have an effect on the difference
between the observed spectral energy distribution (SED) and the one
corresponding to a dustless case \citep[e.g.][]{WG:00,CF:00}.  To
avoid confusion, this difference is termed \textit{attenuation}, instead of
extinction. Nevertheless, the overall behaviour of attenuation in
galaxies is very similar to the extinction law found in stars,
including the presence of the bump
\citep[see, e.g.][]{BBI:05,NPC:09,CSB:10,WCB:11,BNB:12,HF:14,HF:15,RKS:15,SFC:15}.
However, large variations are found among galaxies 
\citep{Cal:99, GSC:99, JSS:07,C:10,KC:13,ZG:15, BCC:16}.
In fact, the standard dust attenuation law used in star-burst
galaxies lacks this bump \citep{Cal:00}, but the latest
radiative transfer models would suggest this law would
need a suppression of the carriers that produce the bump
\citep{SD:16}. There is indeed ample evidence for PAH
destruction/suppression in starbursts relative to normal star-forming
galaxies, \citep[see e.g.][and references therein]{Wu:06, Mu:14},
see also \citet{SKR:16} for evidence of PAH strength variation with other
galaxy properties at high redshift.

Disentangling these complex effects requires robust observational
constraints of the attenuation law in star-forming galaxies.  The
observed variations involve the amount of dust present, grain size
distribution and composition, all affected by the properties of the
stellar populations such as age or metallicity
\citep{GSC:99, PGB:07, SPL:16}.

%%%%%%%%%%%%%%  SSFR   %%%%%%%%%%%%%%%%%
%%%%%%%%%%%%%%%%%%%%%%%%%%%%%%%%%%%%%%%%%%%%%%%%
\begin{figure}
\begin{center}
\includegraphics[width=85mm]{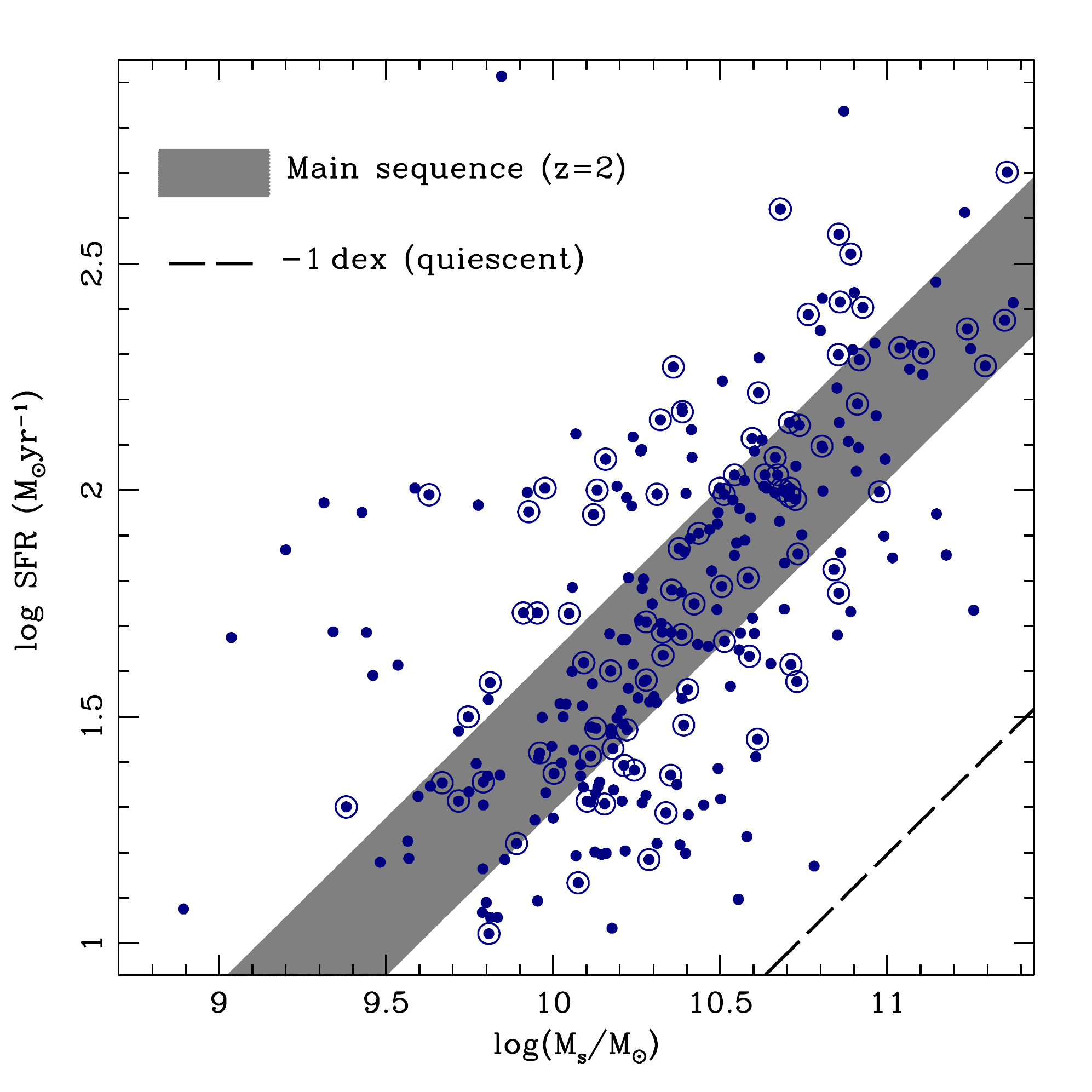}
\end{center}
\caption{Correlation between star formation rate and stellar mass.
Only those sources with available 24\,$\mu$m fluxes are shown (blue
dots). For reference, the Main Sequence at z=2 is shown, as defined
by \citet[][grey shaded region]{Speagle:14}.  The dashed line is an
offset of the above by $-$1\,dex to guide the eye towards the region
where quiescent galaxies should lie. Single dots represent galaxies
with a photometric redshift; whereas those encircled by an outer ring
are galaxies with a spectroscopic redshift. The figure does not
include sources only with an upper limit estimate to the SFR.}
\label{fig:SFR}
\end{figure}
%%%%%%%%%%%%%%%%%%%%%%%%%%%%%%%%%%%%%%%%%%%%%%%%

The goal of this paper is to provide new constraints on the dust
attenuation law of galaxies at higher redshift exploiting the Survey
of High-z Absorption Red and Dead Sources ({\sc SHARDS}).  SHARDS is
an optimal data set for the purpose of this paper since its spectral
resolution, in combination with ancillary data, probes the attenuation
law in the rest-frame NUV and optical windows at redshift z$\sim$2.
Furthermore, the depth of the survey allows us to study
individual galaxies, avoiding stacking and its complications.

The structure of the paper is as follows: in \S\ref{Sec:Data}
the data used in this study are presented. In 
\S\ref{Sec:Method} we describe the methodology, involving a
comparison of the observed photo-spectra with a set of population
synthesis models, including a test with simulated data. The results
are discussed in \S\ref{Sec:Disc}, including an assessment of the
effect of the observed variations in the derivation of the UV
slope, $\beta$. We conclude with a summary of
the results in Section~\ref{Sec:Summ}. 
A standard $\Lambda$CDM cosmology is adopted, with $\Omega_m$=0.27
and H$_0$=72\,km\,s$^{-1}$Mpc$^{-1}$. All magnitudes are quoted in the
AB system, and stellar masses correspond to a \citet{Chabrier:03}
IMF.

%%%%%%%%%%%%%%%%%%%%%%%%%%%%%%%%%%%%%%%%%%%%%%%%
%%%%%%%%%%%%%%%%%  SHARDS   %%%%%%%%%%%%%%%%%%%%
%%%%%%%%%%%%%%%%%%%%%%%%%%%%%%%%%%%%%%%%%%%%%%%%
\begin{figure*}
\begin{center}
\includegraphics[width=160mm]{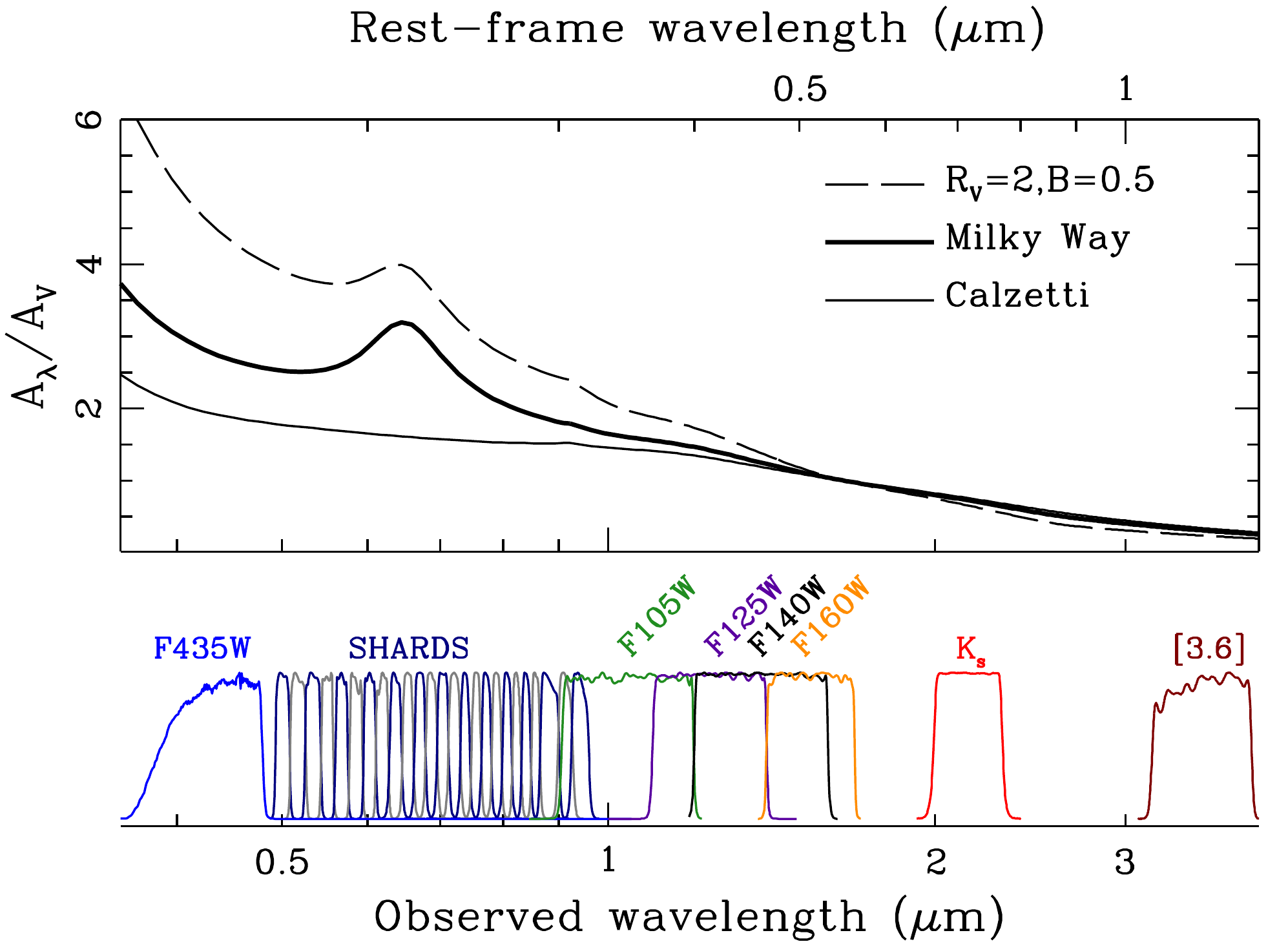}
\end{center}
\caption{The set of filters used in this analysis is shown as
a function of observed (bottom) and rest-frame wavelength (top), with the
latter assuming a galaxy at z=2, roughly corresponding to the median
redshift of the sample. The tranmission curves are normalized at the
peak value. The top panel shows three choices of the 
attenuation law, as labelled.  Note the {\sc SHARDS}
medium band filters sample the region around the NUV 2,175\AA\ bump at
the redshifts of interest (1.5$<$z$<$3).}
\label{fig:attSHARDS}
\end{figure*}
%%%%%%%%%%%%%%%%%%%%%%%%%%%%%%%%%%%%%%%%%%%%%%%%

%%%%%%%%%%%%%%%%%%%%%%%%%%%%%%%%%%%%%%%%%%%%%%%%
\section{Sample definition}
\label{Sec:Data}

The sample is extracted from the Survey of Red and Dead Sources
\citep[{\sc SHARDS},][]{SHARDS}, a data set comprising deep
($<26.5$\,AB at 4\,$\sigma$) optical photometry in a set of 25 medium
band filters (full-width at half-maximum, FWHM$\sim 150$\AA) covering a 
141\,arcmin$^2$ region towards GOODS-N. {\sc SHARDS} effectively provides
low-resolution spectra (R$\sim$50) over the full field, without the
completeness issues of standard multi-object spectroscopy. We
complement the optical photometry with ACS F435W 
\citep{ACS}.
In addition, we use NIR fluxes from CANDELS \citep{CANDELS} in the
{\sl HST}/WFC3 F105W, F125W, F140W and F160W passbands, as well as deep
K$_s$ photometry from CFHT/WIRCam \citep{CFHT} and {\sl Spitzer}/IRAC
3.6\,$\mu$m \citep{IRAC}.  The data are retrieved via the
Rainbow\footnote{\tt http://rainbowx.fis.ucm.es}
database \citep{PG:08,Barro:11}, using version 14.5.  We select
sources with a median S/N per filter of at least 5\,$\sigma$ and with
acceptable GALFIT models in the CANDELS WFC3/F160W images (i.e. no
error flags), as well as with available photometric
redshifts (however note a fraction of our sample has
spectroscopic redshifts, see below). Given the flux constraint in the
NUV-rest frame -- mapped by SHARDS -- we note that this survey is
inherently biased in favour of NUV bright sources.

%%%%%%%%%%%%%%%%%%%%%%%%%%%%%%%%%%%%%%%%%%%%%%%%
%%%%%%%%%%%%  SED examples  %%%%%%%%%%%%%%%%%%%%
%%%%%%%%%%%%%%%%%%%%%%%%%%%%%%%%%%%%%%%%%%%%%%%%
\begin{figure*}
\centering
\includegraphics[width=150mm]{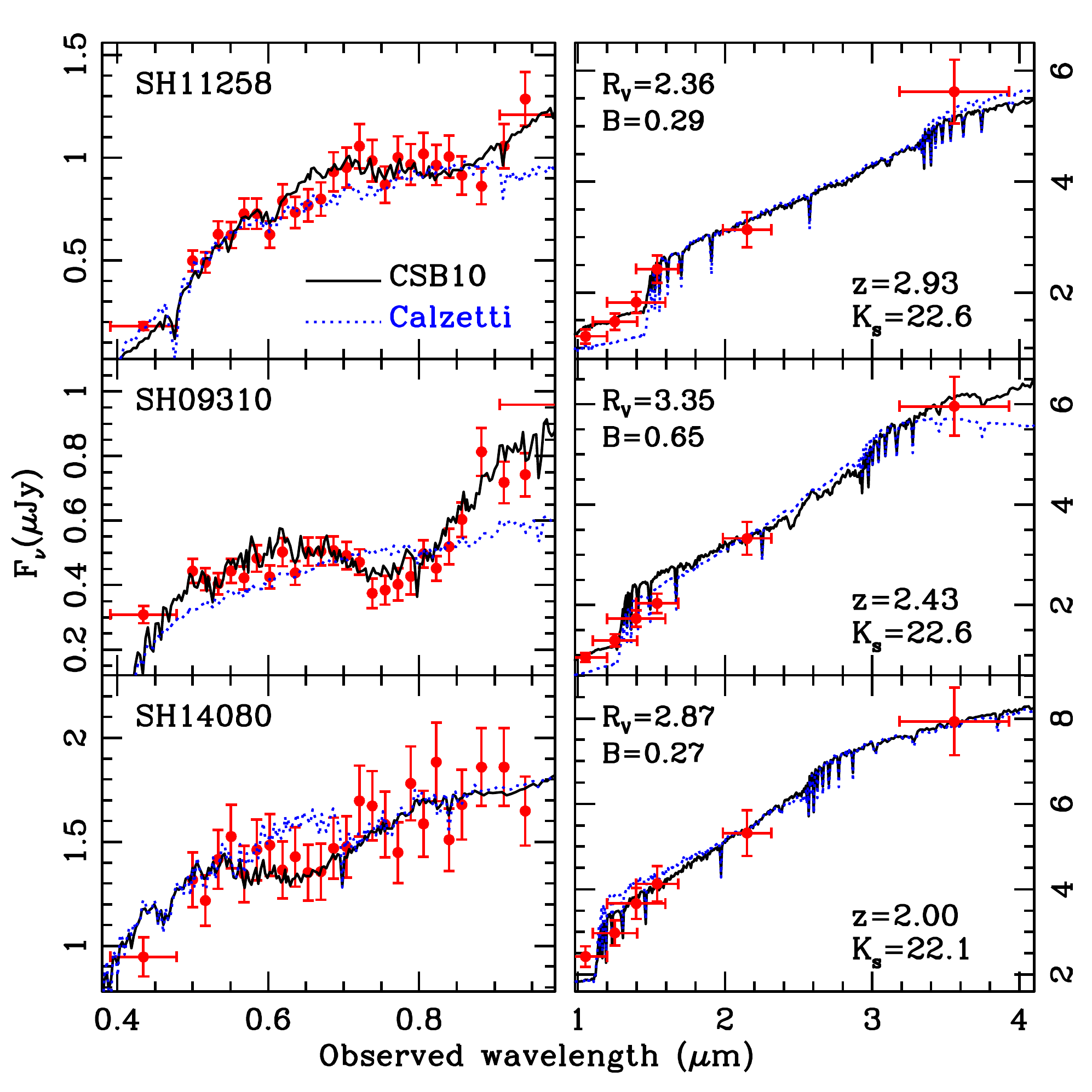}
\caption{Example of the fits to the observed fluxes in three
galaxies, labelled by their SHARDS ID, redshift and K$_s$ AB
magnitude, as well as the best-fit values of B and R$_V$.  The red
dots are the observed fluxes, shown with their flux uncertainties. The
solid line is the best fit to the models, shown at a better resolution
than the photometric data to illustrate the type of populations being
considered. For reference, the best fit model, restricting the
analysis to a Calzetti law is shown as a dotted line. The wavelength
coverage is split between the observed frame optical (left) and NIR
(right). Note the prominent NUV bump in SH09310 (middle-left panel),
mapped at 0.75\,$\mu$m at the redshift of the galaxy.}
\label{fig:fits}
\end{figure*}
%%%%%%%%%%%%%%%%%%%%%%%%%%%%%%%%%%%%%%%%%%%%%%%%

As the project focuses on the dust attenuation law in the rest-frame
NUV + optical window, we select galaxies over a redshift interval such
that the NUV bump is included within the spectral range of {\sc
SHARDS} (0.5--0.9\,$\mu$m). The sample comprises 1,807 galaxies at
1.5$<$z$<$3, with a median redshift of 2.14. We note that 16.7\% of
the sample has available spectroscopic redshifts. This subsample of
302 galaxies has a photo-z accuracy of 0.4 per cent given by the
median of the distribution $|\Delta z|/(1+z)$, mostly thanks to the
photo-spectra provided by the medium-band
filters \citep{SHARDS,Ferreras:14}.  There is no significant trend
between the photometric redshift accuracy and the apparent magnitude
of the sources (between F160W=22 and 25\,AB).  The number of
catastrophic failures -- defined as those photometric redshifts with
an accuracy outside of three times the rms of the distribution --
is 3.9\%. The above criteria favour bright UV sources, therefore, we need to check
whether the photo-spectra are contaminated by AGN light. We
cross-match our sample with the X-ray sources in the 2\,Ms Chandra
Deep Field North \citep{cdfn}, following \citet{Trouille:08} to
convert fluxes in the 2-8\,keV band into luminosities. Only 33 sources
have an X-ray detection, with an average $\langle L_X\rangle =
2.9\times 10^{43}$\,erg\,s$^{-1}$, and only two sources brighter than
$L_X>10^{44}$\,erg\,s$^{-1}$. Nevertheless, we discard all those
sources with an X-ray detection, to make sure our data are dominated
by stellar light. Moreover, we discard potential sources with an
obscured AGN, following the criterion of \citet{Donley:12} based
on flux ratios from Spitzer/IRAC. Our final sample therefore comprises 1,753
star-forming galaxies. Note that one of the adopted selection criteria
requires a good surface brightness fit in F160W, mapping a rest-frame
interval between $B$ and $R$ band.  Furthermore, the distribution of
the S\'ersic index from the CANDELS surface brightness fits in the
WFC3/F160W band (see \S\ref{Sec:Disc}), is rather low, suggesting a
negligible contribution from a central AGN.

Fig.~\ref{fig:SFR} shows the sample with respect to the star formation
rate (SFR), following \citet{Rujopakarn:13}.  In this study, the
authors estimate star formation rates based on single band
measurements with {\sl Spitzer}/MIPS at 24\,$\mu$m,  out to z$<$2.8
(see section~5 of that reference for details).

Note we only show in the figure those
galaxies with an observed 24\,$\mu$m flux, avoiding here galaxies with
just an upper bound to the SFR. For reference, the main sequence at z=2
\citep[as derived by][]{Speagle:14} is shown as a grey shaded region, that
includes the observational scatter in the relation. This redshift is
approximately the median value of our sample. The dashed line
represents the loci of quiescent galaxies by offseting the main
sequence relation by $-$1\,dex in SFR. This figure shows that our
sample lies mainly around the main sequence (grey shaded area) and
covers a wide range of stellar mass. We also note that the sample does
not feature any significant selection bias between mass and redshift.

We follow the prescription proposed by \cite{CSB:10} (hereafter,
CSB10), defined by two parameters, namely the NUV bump strength (B)
and the total to selective extinction ratio (R$_V$).

%%%%%%%%%%%%%%%%%%%%%%%%%%%%%%%%%%%%%%%%%%%%%%%%%%%%%%%%%%%%%%%%%%%%
%%%%%%%%%%%%    Table parameters
%%%%%%%%%%%%%%%%%%%%%%%%%%%%%%%%%%%%%%%%%%%%%%%%%%%%%%%%%%%%%%%%%%%%
\begin{table*}
\caption{Parameter range of the grid of models (see text for details).}
\label{tab:pars}
\begin{center}
%\resizebox{\columnwidth}{!}{%
\begin{tabular}{cccc}
\hline
Observable & Parameter & Range & Steps \\
\hline
Age (log scale) & $\log($t/Gyr$)$ & [-2, +0.6] & 16 \\
Metallicity (log scale) & $\log(Z/Z_\odot)$ & [-2, 0.3] & 8 \\
Colour Excess & E($B-V$) & [0, 1.5] & 24 \\
Total to selective extinction ratio & R$_V$ & [0.5, 5] & 24 \\
NUV Bump strength & B & [0, 1.5] & 24 \\
\hline
  &      & Number of models & 1,769,472\\
\hline
\end{tabular}
%}
\end{center}
\end{table*}
%%%%%%%%%%%%%%%%%%%%%%%%%%%%%%%%%%%%%%%%%%%%%%%%%%%%%%%%%%%%%%%%%%%%

%Table for  corrections to the data 
\begin{table}
\centering
\caption{Corrections to the data: $\pi_{\rm true}=\alpha \pi_{\rm OBS} + \beta$.
Derived from Simulations shown in Fig. ~\ref{fig:allhist}. Col. 1 identifies
the dust attenuation parameter, col. 2 and 3 are the best linear fit parameters
taking all simulated SFHs into account. Col. 4 is the RMS scatter of the difference
between true and extracted values.
}
\label{tab:corr}
\begin{tabular}{lccc}
\hline
$\pi$ & $\alpha$ & $\beta$ & RMS($\Delta\pi$)\\
\hline
R$_V$    & $1.0329$ & $-0.337$ & $0.72$\\
B        & $0.9772$ & $+0.073$ & $0.18$\\
E($B-V$) & $0.7738$ & $+0.032$ & $0.07$\\
\hline
\end{tabular}
\end{table}

%%%%%%%%%%%%%%%%%%%%%%%%%%%%%%%%%%%%%%%%%%%%%%%%
\section{Constraining the dust attenuation law}
\label{Sec:Method}

Following the aforementioned CSB10 dust attenuation law at the
fiducial value B=1, R$_V$=3.1, this parameterisation recovers the
standard Milky Way extinction law of \citet{CCM:89}. A third parameter
controls the net amount of reddening, and can be described either by
the attenuation in a reference band (e.g. A$_V$) or by a colour excess
(e.g. E($B-V$)).  We adopt the latter.

Fig.~\ref{fig:attSHARDS} shows the transmission curves of the 32
filters used in this study, including the {\sl HST}/ACS $B$-band, 25 SHARDS
filters, the NIR {\sl HST}/WFC3 passbands from CANDELS, CFHT/WIRCam K$_s$
and {\sl Spitzer}/IRAC 3.6\,$\mu$m. This set spans a wide enough spectral range
to constrain the dust-related parameters. The SHARDS fluxes provide a good sampling of
the region around the NUV bump over the selected redshift range. For reference,
the figure includes the \citet{Cal:00} law (thin line), the
standard Milky Way extinction (thick line), 
and an additional case from CSB10 (dashed line), adopting a weaker bump (B=0.5) and
a steeper wavelength dependence (R$_V$=2).
This paper does not aim at exploring the distribution of dust within
galaxies. Therefore, we only consider the effect of dust as a
foreground screen acting on the underlying stellar light.  Such a
simplied model gives a robust constraint on the attenuation
law. The only difference between the assumption of a foreground screen or
a homogeneous distribution of dust and stars is in the apparent optical depth.
The drawback is, of course, the inability to disentangle dust
properties from dust-star geometry in the interpretation of the
results. Our aim is, therefore, to use the observed fluxes to constrain
the attenuation law, which requires marginalizing over the parameters
that control the illumination source (i.e. the stellar populations).
We use the stellar population synthesis models of
\citet[][hereafter BC03]{BC03}, with the standard assumption of a
\citet{Chabrier:03} initial mass function.
The adopted range of parameters can be found in Table~\ref{tab:pars}. 

%%%%%%%%%%%%%%%%%%%%%%%%%%%%%%%%%%%%%%%%%%%%%%%%
%%%%%%%%%%%%%%%  Simulations   %%%%%%%%%%%%%%%%%
%%%%%%%%%%%%%%%%%%%%%%%%%%%%%%%%%%%%%%%%%%%%%%%%
\begin{figure}
\begin{center}
\includegraphics[width=85mm]{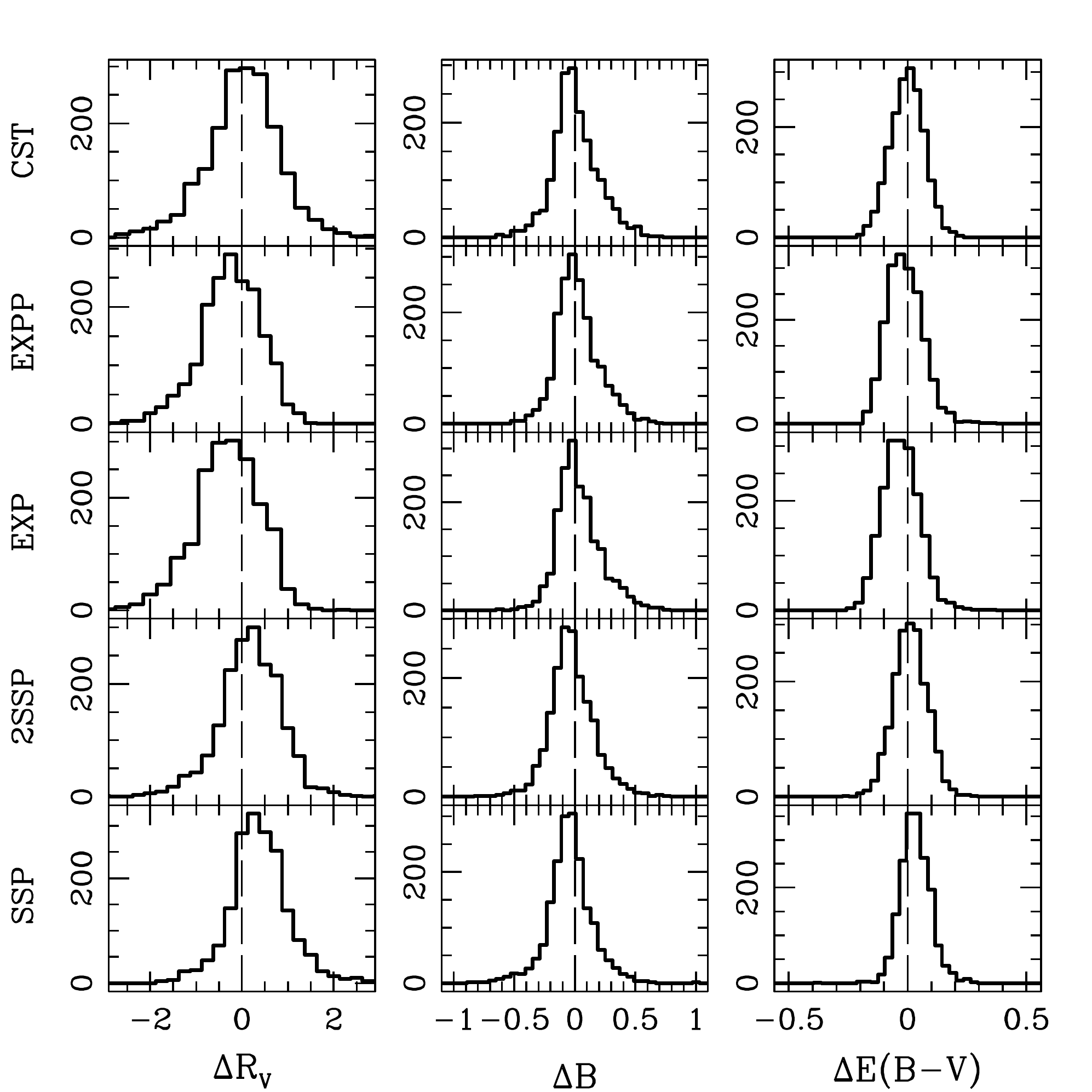}
\end{center}
\caption{ 
We illustrate the accuracy of the method in constraining the dust
attenuation parameters by comparing input and retrieved values in a
set of simulations ($\Delta$R$_V$, $\Delta$B and
$\Delta$E($B-V$)). Five sets of SFHs are considered, from top to
bottom: constant star formation rate (CST)); exponentially increasing 
star formation rate (EXPP); exponentially decaying star formation
rate (EXP); a superposition of two simple stellar populations (2SSP)
and a simple stellar population (SSP). The width of the distributions
provide an estimate of the uncertainties expected in the analysis (see
col. 4 in Table~\ref{tab:corr}).}
\label{fig:allhist}
\end{figure}
%%%%%%%%%%%%%%%%%%%%%%%%%%%%%%%%%%%%%%%%%%%%%%%%

Our methodology entails a comparison between a large grid of BC03
models and the observations. The grid assumes simple stellar
populations (SSP), corresponding to a single age and metallicity. We
define a large set of lookup tables of model photometry to be used
with all galaxies. Therefore, redshift is an additional parameter in
the grid. An important issue to bear in mind is that the response
profile of the medium band filters of the SHARDS survey depends on the
position of the source on the focal plane \citep{SHARDS}. The net
effect is a position-dependent shift of the transmission curve in
wavelength, and this shift can be as large as the extent (FWHM) of the
passband \citep[see][]{SHARDS,Cava:15}. Therefore, each source
effectively \textit{sees} an independent set of SHARDS filters, in
principle requiring an individual grid of model fluxes accounting for
those filter offsets, prohibitively expensive in computing time.
However, this effect can be modelled accurately by a linear fit of the
model fluxes to the central wavelength of each filter.  Since the
2,175\AA\ bump is the only remarkable spectral feature in the range
covered by SHARDS at these redshifts, and it is much wider than the
FWHM of the filters, a simple linear parameterisation of the effect of
these shifts on the photometry gives accurate results, with residuals
well below the flux uncertainties. We implement this approximation for
each of the 25 SHARDS filters by measuring the model fluxes shifted by
$\pm 100$\AA\ with respect to the central wavelength, in order to
create two different sets of grids from which we define the linear
interpolation. Regarding redshift, we create one set of grids at
intervals $\Delta$z=0.1 between z=1.5 and z=3. For each galaxy we
apply a linear interpolation between the reference redshift values
from the grid that straddle the observed redshift.

%%%%%%%%%%%%%%%%%%%%%%%%%%%%%%%%%%%%%%%%%%%%%%%%
%%%%%%%%%%%%%%%%  Nebular   %%%%%%%%%%%%%%%%%%%%
%%%%%%%%%%%%%%%%%%%%%%%%%%%%%%%%%%%%%%%%%%%%%%%%
\begin{figure}
\centering
\includegraphics[width=85mm]{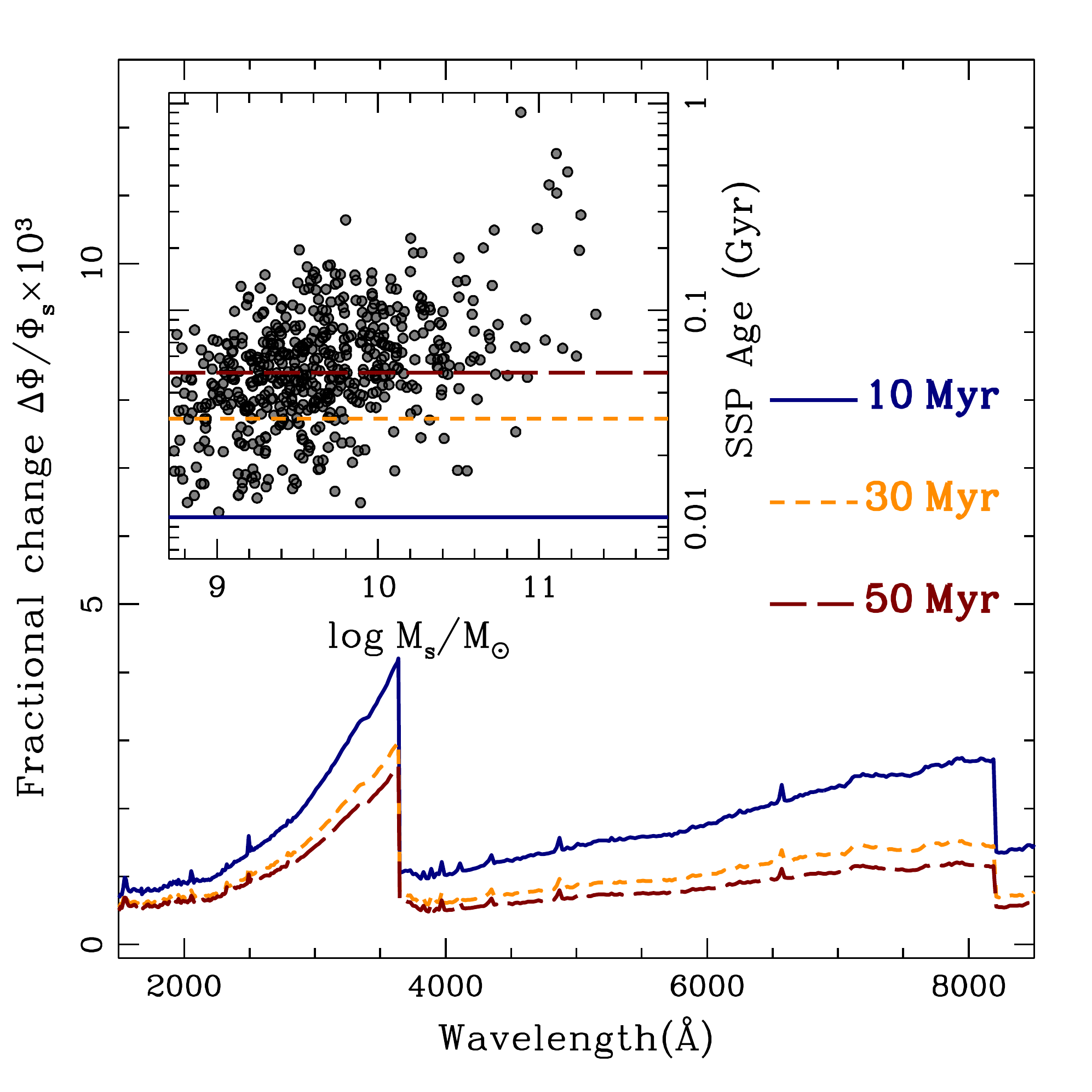}
\caption{Relative change in the flux contributed by
nebular continuum emission at three different ages, following the {\sc
STARBURST\,99} models \citep{SB99} at solar metallicity, for a
continuous star formation history with a Salpeter IMF. The inset shows the
derived {\sl SSP equivalent} ages of our sample, following the same colour coding of
the three choices of SSP age. To avoid overcrowding the plot, only one
out of three data points (randomly chosen) are shown.}
\label{fig:nebular}
\end{figure}
%%%%%%%%%%%%%%%%%%%%%%%%%%%%%%%%%%%%%%%%%%%%%%%%

We follow a standard Bayesian approach, so that the expectation
value of each of the parameters is given by the mean of the
marginalized probability distribution function, assuming flat priors
on the parameters, over a range shown in Table ~\ref{tab:pars}, and a
likelihood function defined by a $\chi^2$ statistic:
\begin{equation}
{\cal L}(\pi_i)\propto \exp\Bigg\{-\frac{1}{2}
\sum_{j}\Bigg(\frac{\Phi_{j}^{OBS} - \Phi_{j}^{MODEL} (\pi_{i})}{\sigma(\Phi_{j})}\Bigg)^2
\Bigg\},
\label{eq:chi2eq}
\end{equation}
where $\Phi_{j}^{OBS}$ and $\Phi_{j}^{MODEL}$ are the observed and
model fluxes, respectively. $\sigma(\Phi_{j})$ corresponds to the
(1\,$\sigma$) photometric uncertainties. $\{\pi_{i}\}$ represent the
model parameters, namely five: SSP age and metallicity, along with
B, R$_V$ and E($B-V$).

Fig.~\ref{fig:fits} illustrates the typical fitting results for three
galaxies at K$_s\!\sim$22-23\,AB, where the observed fluxes are
plotted in red, and the best-fit model as a solid black line.  The
figure is split left/right between the optical and the NIR windows. We
include the FWHM of the NIR passbands as horizontal error bars, and
show the best fit at a higher spectral resolution than the one used in
the (photometric) analysis to show the type of stellar populations
typically found in our sample, with SSP-equivalent ages between 10\,Myr
and 1\,Gyr (see inset in Fig.~\ref{fig:nebular}). Each galaxy is
labelled by their SHARDS ID, redshift and magnitude in the K$_s$ band,
and the best-fit to the dust parameters B and R$_V$ is also shown.  In
general, the whole sample produces good fits, with an average, reduced
$\langle\chi^2_r\rangle= 1.54$ (considering 32 fluxes and 5 free
parameters). The blue dotted lines represent the best-fit case when
enforcing a \citet{Cal:00} attenuation. Note in all cases the fit is
significantly worse for this choice of attenuation.  Note the
prominent NUV bump in the SHARDS photometry of galaxy SH09310 at
0.75\,$\mu$m in the middle-left panel, for which a Calzetti law cannot
give a good fit.

There are two main issues that need to be assessed regarding the
methodology: 1) How well does the grid-based method behave with
respect to the linear interpolations performed both in redshift and
within the transmission curve of each of the 25 SHARDS filters?  2) We
approximate the illumination source as a simple stellar
population. How well can this treatment reproduce the complex
populations typically found in galaxies? These can be addressed by
running a set of simulations, comparing the input values of the dust
attenuation law with the retrieved values following this
methodology. We create five sets of mock data, each corresponding to a
different parameterisation of the star formation history: A simple
stellar population (SSP), representing single burst galaxies; a
two-burst model (2SSP) that overlays a young component over an older
one, an exponentially decaying rate (EXP), an exponentially increasing
rate (EXPP) and a constant star formation rate followed by a
truncation (CST). These sets of synthetic data represent the typical
range of potential formation histories explored in the literature
\citep[e.g.][]{Papovich:11,Reddy:12,FW4871,HDS:16}. In all cases the
illumination source derived from the mixture of stellar populations is
subject to a single foreground screen with parameters B, R$_V$, and
E($B-V$). Each of the five sets comprise the same number of galaxies
as the original sample, with the same redshift distribution and
photometric uncertainties. For each simulated galaxy, we draw a random
estimate for each of the parameters that define the sample
\citep[see][for details]{HF:15}, therefore we assume the mock sample
has uncorrelated dust parameters. This is a property we exploit below to
assess potential systematics.

The difference between the \textit{true} input values and those retrieved
from the SSP grid-based analysis is presented in
Fig.~\ref{fig:allhist}, for the five different choices of star
formation history. The mock data explore a wide range of the age
distribution, metallicity and dust. These figures show the comparison
between the input and the retrieved values. Even though the actual star
formation histories of galaxies are more complex than the SSP-adopted
set in our grids, this exercise illustrates how well one can recover
the attenuation parameters by marginalising over a large grid of SSPs.
A similar result, although based on a different set of photometric filters,
was found by \citet{HF:15}. The simulations can also be used to
\textit{calibrate} the methods, by performing a linear fit between the
input and the output parameters. We note that this calibration step is
not essential, but it helps to increase the accuracy in the
determination of the dust-related parameters. Table ~\ref{tab:corr}
shows the calibration fits. In order to avoid the effect of outliers,
the fits are obtained after clipping the sample at the 3\,$\sigma$ level.
The table also shows the RMS scatter of the
comparison between the input value and the retrieved value, which
serves as an estimate of the accuracy of the method, namely
$\Delta$B=0.18; $\Delta$R$_V$=0.75 and $\Delta$E($B-V$)=0.10\,mag.

%%%%%%%%%%%%%%%%%%%%%%%%%%%%%%%%%%%%%%%%%%%%%%%%
\subsection{Effect of nebular emission}
\label{SSec:Nebular}

The most significant contribution from nebular emission is in
relatively narrow spectral lines. Our methodology removes outliers to
the fits at specific wavelengths, so we are resilient to such
contamination. However, an additional contribution that can affect the
flux measured in all passbands is from the nebular continuum. Since
emission from the nebular continuum is not included in BC03 models, we
explore the impact of that effect in our method using a set of simple
stellar populations from the {\sc STARBURST\,99} models \citep{SB99}.
We assume solar metallicity and a Salpeter IMF for practicality, and
compare this relative change for three choices of the age: 10\,Myr
(solid blue); 30\,Myr (short dashed orange) and 50\,Myr (long dashed
red), within the spectral range used here.  Fig.~\ref{fig:nebular}
justifies our approximation by comparing the relative flux increase
contributed by nebular emission. We show three cases corresponding to
young populations from the Starburst99 models \citep{SB99}, adopting a
constant star formation history, Salpeter IMF, and solar metallicity.
Even at the youngest ages ($\sim$10\,Myr), the contribution from the nebular
continuum stays below the $\sim$0.5\,\% level. For reference, the inset
in Fig.~\ref{fig:nebular} shows the retrieved SSP ages from the sample
with respect to stellar mass. Note that the majority ($\sim$81\%) of
our sources have SSP ages older than 30\,Myr. Since an uncorrected SSP
age should be {\sl younger} if nebular emission is not accounted for,
we conclude that this component is not essential in our analysis.

%%%%%%%%%%%%%%%%%%%%%%%%%%%%%%%%%%%%%%%%%%%%%%%%
\section{Discussion}
\label{Sec:Disc}

%%%%%%%%%%%%%%%%%%%%%%%%%%%%%%%%%%%%%%%%%%%%%%%%
%%%%%%%%%%%%%%%%%  SHARDS   %%%%%%%%%%%%%%%%%%%%
%%%%%%%%%%%%%%%%%%%%%%%%%%%%%%%%%%%%%%%%%%%%%%%%
\begin{figure}
\centering
\includegraphics[width=85mm]{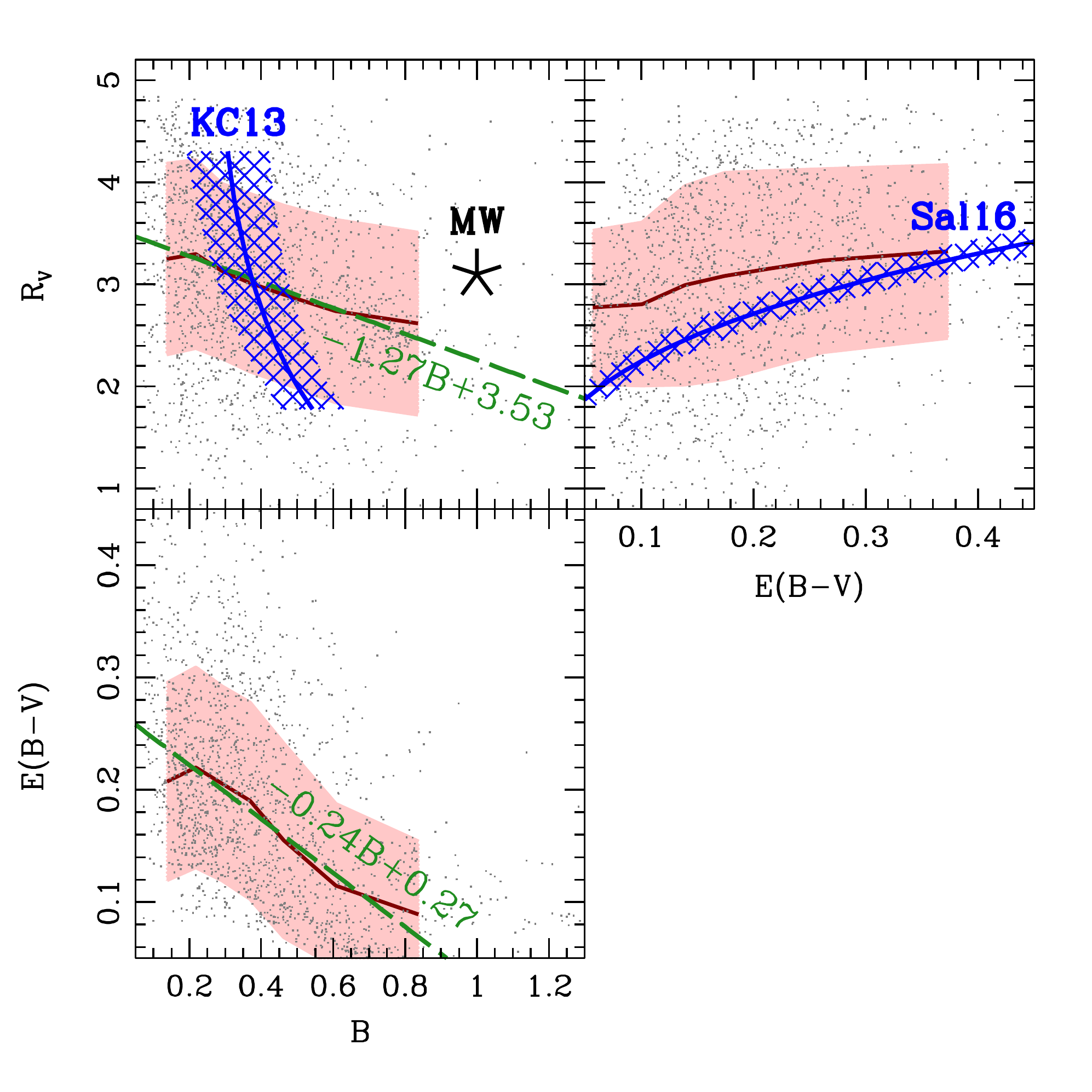}
\caption{Correlations between dust attenuation parameters.
Individual data points are shown as grey dots, whereas the thick solid
line and shading represent a moving median and RMS scatter,
respectively. The star in the top-left panel (labelled MW) corresponds
to the standard extinction law in the Milky Way, and the blue hatched
region mark the estimates from \citet[][labelled KC13]{KC:13}.  The top-right panel
shows, also in blue, the trend presented in \citet[][labelled Sal16]{SPL:16}. These two
comparisons required the conversion from their attenuation slope
$\delta$ to a standard R$_V$ (see Appendix~\ref{KCeq} for details).
We also present a simple linear fit to the correlations on the left
panels (green dashed lines), described by equations~\ref{eq:dust1}
and \ref{eq:dust2}.  Note we do not attempt such a fit between R$_V$
and E($B-V$) because there is a large scatter of the data
points. On \citet{KC:13}, both sSFR and stellar mass are derived
galaxy properties, with the equivalent width of H$_{\alpha}$ is used
as a SFR proxy.  }
\label{fig:parpar}
\end{figure}
%%%%%%%%%%%%%%%%%%%%%%%%%%%%%%%%%%%%%%%%%%%%%%%%

%%%%%%%%%%%%%%%%%%%%%%%%%%%%%%%%%%%%%%%%%%%%%
\begin{figure*}
\centering
\includegraphics[width=180mm]{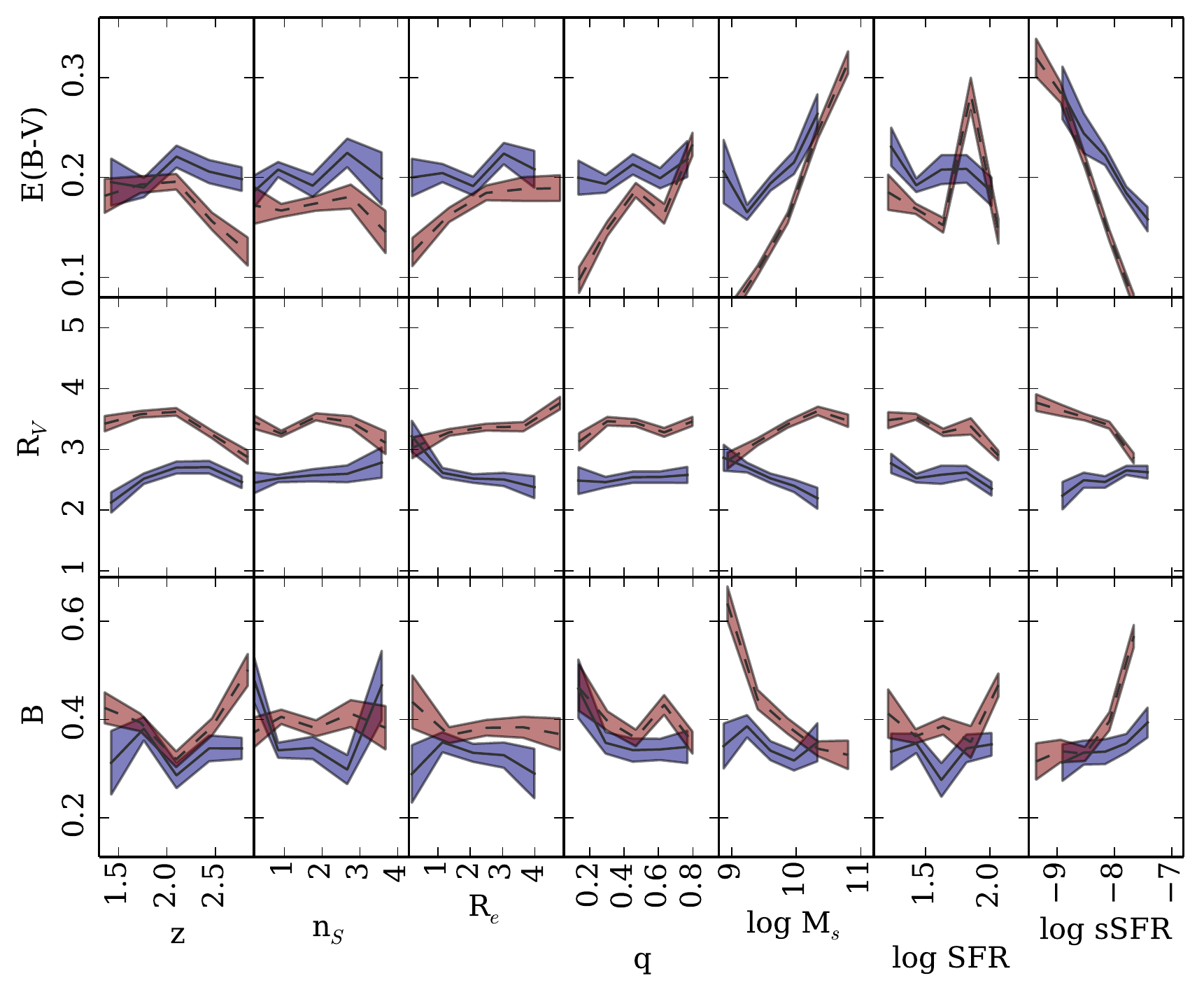}
\caption{Relation between the dust-related parameters of the dust attenuation law
(from top to bottom: colour excess, total to selective ratio and NUV
bump strength) and several observables, from left to right: spectroscopic
redshift, S\'ersic index, semi-major axis in physical units (kpc) --
measured in the WFC3/F160W band, axis ratio, stellar mass (in M$_\odot$), star formation
rate (in M$_\odot$\,yr$^{-1}$) and specific star formation rate (in yr$^{-1}$).
In all panels the lines (and
shade) trace the median (and 1\,$\sigma$ error) of subsamples, split
with respect to stellar age (blue-solid: young tercile, red-dashed: old
tercile). The median stellar age of the distribution is 5.2\,Myr. From the plots displayed above, we observe that the most significant correlation is seen between the specific star formation rate and two of the dust parameters, the colour excess and NUV bump strength. In particular, they both exhibit oppposite trends. An increment in sSFR correlates with a decrease in colour excess and is associated with a stronger NUV bump strength.  
Furthermore, the trend of the dust-related parameters with stellar mass is shown in Table~\ref{tab:lMs}.}
\label{fig:Pars}
\end{figure*}
%%%%%%%%%%%%%%%%%%%%%%%%%%%%%%%%%%%%%%%%%%%%%

%%%%%%%%%%%%%%%%%%%%%%%%%%%%%%%%%%%%%%%%%%%%%%%%%%%%%%%%%%%%%%%%%%%%
%%%%%%%%%%%%    Table relation with stellar mass
%%%%%%%%%%%%%%%%%%%%%%%%%%%%%%%%%%%%%%%%%%%%%%%%%%%%%%%%%%%%%%%%%%%%
\begin{table*}
\caption{Correlation between stellar mass and dust-related
parameters, as shown in Fig.~\ref{fig:Pars}. Each column gives the
median value and the RMS scatter per bin in stellar mass. The
results are shown for the younger and older terciles of the distribution.
The last column (n) gives the number of data points per bin.}
\label{tab:lMs}
\begin{center}
%\resizebox{\columnwidth}{!}{%
\begin{tabular}{ccccc}
\hline
%Q1/Median/Q4: 0.034 0.0521 0.0754
 log M$_{\rm s}$/M$_\odot$ & E(B-V) &  R$_V$ &  B & n\\
\hline
\multicolumn{4}{c}{YOUNG (t$_{\rm SSP}\simlt$\,4.1\,Myr)}\\
\hline
   8.69-- 9.06 & 0.21$\pm$0.15 &  2.86$\pm$1.03 &  0.35$\pm$0.22 &   24\\
   9.06-- 9.42 & 0.17$\pm$0.10 &  2.70$\pm$1.08 &  0.39$\pm$0.30 &  179\\
   9.42-- 9.78 & 0.19$\pm$0.09 &  2.52$\pm$0.93 &  0.34$\pm$0.23 &  138\\
   9.78--10.15 & 0.21$\pm$0.10 &  2.40$\pm$0.86 &  0.32$\pm$0.17 &   71\\
  10.15--10.51 & 0.26$\pm$0.11 &  2.20$\pm$0.95 &  0.35$\pm$0.21 &   31\\
\hline
\multicolumn{4}{c}{OLD (t$_{\rm SSP}\simgt$\,6.7\,Myr)}\\
\hline
   8.70-- 9.17 & 0.06$\pm$0.05 &  2.81$\pm$0.72 &  0.64$\pm$0.19 &   30\\
   9.17-- 9.63 & 0.11$\pm$0.06 &  3.12$\pm$0.77 &  0.44$\pm$0.25 &  162\\
   9.63--10.10 & 0.16$\pm$0.08 &  3.41$\pm$0.80 &  0.39$\pm$0.22 &  193\\
  10.10--10.57 & 0.25$\pm$0.07 &  3.63$\pm$0.73 &  0.34$\pm$0.16 &  105\\
  10.57--11.04 & 0.32$\pm$0.09 &  3.47$\pm$0.80 &  0.33$\pm$0.23 &   62\\
\hline
\end{tabular}
%}
\end{center}
\end{table*}
%%%%%%%%%%%%%%%%%%%%%%%%%%%%%%%%%%%%%%%%%%%%%%%%%%%%%%%%%%%%%%%%%%%%

It is relevant for our purposes to look for potential correlations
between dust attenuation parameters, as shown in
Fig.~\ref{fig:parpar}. The individual data are plotted as grey dots,
whereas the solid red line and shaded region trace the running median
and the corresponding RMS scatter, respectively.  We observe a
decreasing trend between R$_{V}$ and $B$, with a least squares linear
fit given by the green dashed line, described by:
\begin{equation}
R_V = -1.27\,B + 3.53 \qquad ({\rm RMS}=0.89).
\label{eq:dust1}
\end{equation}
It is interesting to note that in the limit $B\rightarrow 0$ we obtain
R$_V=3.53$, i.e. slightly greyer than the standard Milky Way reference, and
comparable with the \citet{Cal:00} law, R$_V=4.05\pm0.80$.  A
qualitatively similar trend is observed by \citet{KC:13}. However,
note that they use a slightly different parameterisation of the
attenuation law -- following \citet{NPC:09} -- adopting a power law
behaviour with index $\delta$ that roughly maps R$_V$, and a bump
strength E$_b$, closely related to our $B$ (see Appendix ~\ref{KCeq}).
The relationship from equation~3 of \cite{KC:13} is shown in
Fig.~\ref{fig:parpar} as a thick blue line, including a hatched region
that accounts for a $\Delta$E$_b=\pm 0.5$ scatter in the relation.
Note that our relationship is qualitatively in agreement with theirs,
although our sample is extended over a much wider range of bump
strength parameters. We also note that, in contrast, our sample gives
constraints on the attenuation law for {\sl individual} galaxies, not
for composite SEDs obtained by stacking a wide range of
systems. Furthermore, our sample has a significant overlap with the
Milky Way standard, represent by a large star symbol (cf. fig.~2
of \citealt{KC:13}).

The bottom-left panel of Fig.~\ref{fig:parpar} suggests a decreasing
colour excess when the attenuation law has a strong NUV bump,
following a linear trend given by the green dashed line, described by:
\begin{equation}
{\rm E}(B-V) = -0.24\,B + 0.27  \qquad ({\rm RMS}=0.08).
\label{eq:dust2}
\end{equation}
Note the RMS of the residual with respect to the fits is comparable to
the uncertainties in the retrieval of the parameters
(Table~\ref{tab:corr}). We make use of the Spearman rank correlation
in order to address possible degeneracies in our dust-related
variables.  The Spearman rank correlation between B and R$_V$ is
$\rho=-0.27$, in contrast with the null hypothesis of no correlation,
obtained from $10^4$ randomized realizations of the same set,
$\rho_{\rm STAT}=0.00\pm 0.02$. To assess a potential systematic trend
from the methodology, we use the simulations presented
in \S\ref{Sec:Method}, that, by construction, adopt a random
distribution, i.e. no correlation between the input parameters.  The
Spearman rank correlation for the output parameters of these mock data
is $\rho_{\rm SYS}=-0.07$, which implies a slight systematic, but not
strong enough to explain the observed trend.
An equivalent comparison between B and E($B-V$) gives $\rho=-0.58$
for the observed correlation, $\rho_{\rm STAT}=0.00\pm 0.02$ for the statistical
expectation with a random sample, and $\rho_{\rm SYS}=-0.04$ for the
potential systematic from the methodology.

The top-right panel compares R$_V$ and E($B-V$), showing an overall
wide range of values. Although the running median suggests an increase
of R$_V$ with colour excess, we decide not to quote any linear fit
given the large scatter found in the individual data points.
\citet{SPL:16} also relate the total-to-selective ratio and
reddening with $\delta$ and E$_{b}$, finding that galaxies with a
higher dust content obey a flatter attenuation curve (i.e. higher
R$_V$), qualitatively following the trend of our running median, as
shown by the blue hatched region in the top-right panel -- a similar
conversion from $\delta$ to R$_V$ was performed, following a least
squares fit (Appendix~\ref{KCeq}).  We note that our results
qualitatively follow the prediction of \citet{Chevallard:13}, who
suggest a trend towards a
steeper (i.e. lower R$_V$) attenuation law at lower opacities,
i.e. lower E($B-V$). However, the scatter in our results is too large
to derive any trend between these two parameters.

The observed trend between R$_V$ and $B$ could have two possible
origins. First, we may assume that the observed variations are due to
changes in the composition of the dust. In this case, we would expect
that very small dust particles are responsible for the bump, as
commonly accepted \citep[e.g.][]{Draine:89}. Small grains will produce
a steeper attenuation law, closer to Rayleigh scattering, therefore
leading to low values of R$_V$, and -- according to the observations
-- to a strong NUV bump. Alternatively, different dust geometries may
play an important part, due to the fact that the attenuation law can
be age-dependent within the complex distribution of stellar
populations in galaxies.  Radiative transfer models show that a
patchier distribution of dust within the galaxy will produce {\sl
both} a weaker bump and a greyer (i.e. higher R$_V$)
attenuation \citep[e.g.][]{WG:00,CF:00,PGB:07}.  Therefore, the
observed correlation, described by eq.~\ref{eq:dust1} could be
interpreted as a trend with respect to the clumpiness of the dust
distribution. We also find that a stronger bump is found at lower
reddening. Note that the colour excess stays below 0.5\,mag.  This is
possibly due to a selection bias since dustier galaxies would appear
fainter than the flux limit of the SHARDS survey.

The relation between the dust-related parameters and general galaxy
properties is presented in Fig.~\ref{fig:Pars}, showing, from left to
right, redshift, the S\'ersic index of the CANDELS surface brightness
fits in the WFC3/F160W band, its corresponding effective radius (in
kpc, physical units) and axis ratio (q), the stellar mass (in units
of M$_\odot$), star formation rate (SFR, in M$_\odot$\,yr$^{-1}$),
and specific star formation rate (sSFR, in yr$^{-1}$). The lines
follow a running median, split with respect to stellar age, with
red-dashed (blue-solid) representing the oldest (youngest) tercile of
the age distribution.  The shaded regions mark the error in the
median.  In general, there are no strong trends with these parameters,
except for a consistent split with respect to stellar age.  Note the
steeper trend for the older population with respect to stellar mass
(i.e. an increasing dust content, with a weaker bump in
more massive galaxies), mirrored
in the trend with respect to sSFR.  This correlation is not
unexpected \citep[e.g.][]{ZYKK:13}, since dust is a component
associated with the presence of star-forming gas.  Regarding
differences with respect to stellar age, the largest offset is
found with respect to R$_V$. For instance, at fixed stellar mass,
older galaxies will have greyer extinction (i.e. higher R$_V$).  No
significant correlation is found between the NUV bump strength and most 
of the parameters shown in Fig.~\ref{fig:Pars}, although the younger
populations consistently feature slightly weaker bumps
\citep[cf.][]{KC:13, WCB:11}. For reference, Table~\ref{tab:lMs}
lists the correlation between stellar mass and the dust-related
parameters.

Note the method is potentially degenerate between R$_V$ and age: a
steep i.e. lower R$_V$ will remove more UV/blue photons, an effect
that is mimicked by changing the age to older values. Therefore, the
potential systematic goes in the {\sl opposite direction},
strengthening the observed result as an intrinsic trend. In addition,
metallicity may also contribute to this degeneracy, but the
observational constraint on metallicity with the available data is
very poor.

%%%%%%%%%%%%%%%%%%%%%%%%%%%%%%%%%%%%%%%%%%%%%%%%
%%%%%%%%%%%%%  Beta correction   %%%%%%%%%%%%%%%
%%%%%%%%%%%%%%%%%%%%%%%%%%%%%%%%%%%%%%%%%%%%%%%%
\begin{figure}
\centering
\includegraphics[width=85mm]{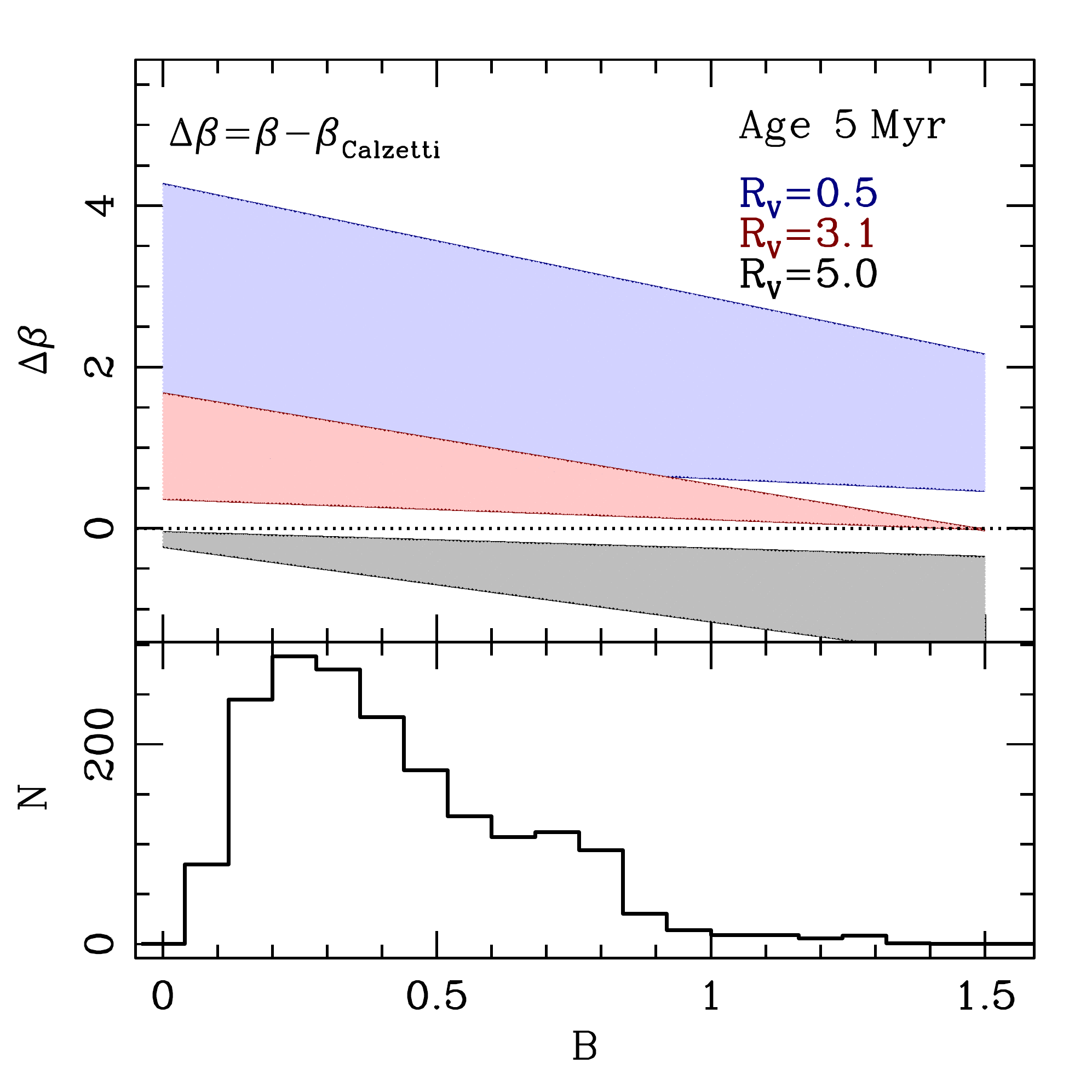}
\caption{Top: Estimate from a synthetic population of the
difference ($\Delta\beta$) between the UV-continuum slope adopting a 
\citet[][CSB10]{CSB:10} dust parametrisation and a 
Calzetti law (R$_V$ = 4.05 and B=0). The model assumes a 5\,Myr
population with solar metallicity, and explores a range of bump
strength values ($B$, horizontal axis). The dotted line indicates
$\Delta\beta = 0$. The different shaded regions represent three
choices of R$_V$, from top to bottom \{0.5,3.1,5.0\}, each region
ranges between E(B-V)=0.1 and 0.5\,mag.  Bottom: Distribution of
the observed NUV bump strength parameters in our SHARDS sample.}
\label{fig:dbeta}
\end{figure}
%%%%%%%%%%%%%%%%%%%%%%%%%%%%%%%%%%%%%%%%%%%%%%%%

\subsection{Variations in the attenuation law and the NUV slope}
\label{SSec:beta}

The results presented above show a significant variation of the dust
attenuation parameters among galaxies (Fig.~\ref{fig:parpar}).  We
consider now the effect of this variation on the derivation of the UV
power law index, $\beta$, used as a standard proxy of dust content in
galaxy spectra. Although the standard definition of
$\beta$ \citep{Cal:94} purposefully avoids a region close to the NUV
bump, this feature is rather broad, so it is important to assess the
effect of these variations.  Moreover, since the NUV bump strength
appears to be correlated with other galaxy properties, one could
expect a systematic bias in the measurement of $\beta$.

In order to quantify this effect, we construct a set of synthetic
populations -- from the \citet{BC03} models -- subjecting them to
various levels of dust attenuation.  Fig.~\ref{fig:dbeta} shows an
example for a 5\,Myr simple stellar population at solar metallicity,
affected by a range of attenuation laws.  We compare the extracted
$\beta$ between a standard
\citet{Cal:00} law and a generic CSB10 law for several choices of
R$_V$ (as labelled). 

The slope $\beta$
is derived following the standard procedure, fitting a power law to
the flux in the NUV region $F(\lambda)\propto \lambda^\beta$, over the
1,300--2,600\AA\ spectral window defined in \citet{Cal:94}.

We define $\Delta\beta$ as the difference between the value from the
CSB10 attenuation case, and the standard (i.e. Calzetti) choice, {\sl
for the same stellar population parameters and with the same colour
excess E($B-V$).}  We consider a range of NUV bump strengths as given
by the horizontal axis. The bottom panel shows the actual distribution
of best-fit B from the SHARDS dataset. The shaded areas in the top
panel extend over a [0.1-0.5]\,mag range of colour excess, E($B-V$).  We
note that if we choose as reference the standard Milky Way extinction
law \citep{CCM:89}, the results do not change appreciably. A robust
determination of UV slope is only possible with young populations
($\simlt$200\,Myr), where the spectrum in the UV region is well fit by
a power law. The chosen stellar age, or metallicity, does not affect
$\Delta\beta$, as long as the population is young enough (i.e. within
the aforementioned range).

%Histogram beta
\begin{figure}
\centering
\includegraphics[width=85mm]{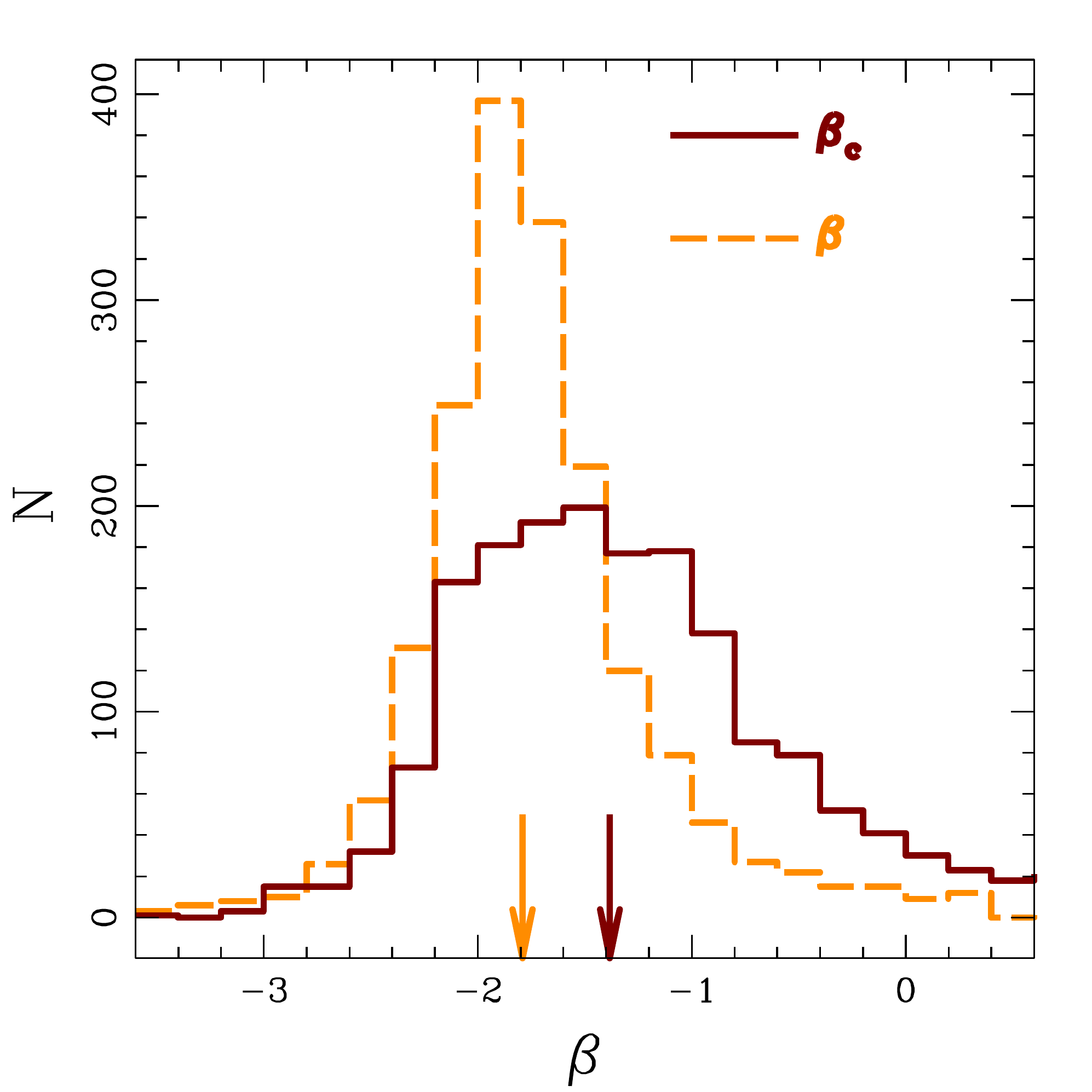}
\caption{Histogram of the distribution of the original measurements
of the UV slope, $\beta$, from the {\sc SHARDS} data (orange dashed
line) and the corrected slope, $\beta_c$ (red solid line, see
Fig.~\ref{fig:dbeta} and text for details).  The arrows mark the
median of both distributions.  }
\label{fig:histobeta}
\end{figure}

Fig.~\ref{fig:dbeta} shows that the UV slope correction can
potentially lie in the range $\Delta\beta\in [-1,4]$, 
depending on the dust-related parameters. To assess
the effect of the variation in dust attenuation properties among
star-forming galaxies, we decided to estimate the UV slope for the
{\sc SHARDS} sample ($\beta_{\rm obs}$), and to correct it to a common
reference -- adopting the \citet{Cal:00} law -- by taking into account
the constraints on the attenuation parameters, on a galaxy-by-galaxy
basis.  Therefore, we define a corrected UV slope:
$\beta_c\equiv \beta_{\rm obs} - \Delta\beta$.
Quantitative estimates on the effects of variations in the NUV bump
and R$_V$ on $\beta$ have been discussed previously\citep[see,
e.g.,][]{BGH:11,BNB:12,RKS:15}. 

We obtained the slope following the procedure mentioned above, using the
SHARDS photometric measurements corresponding to filters that fall
within the intervals proposed in \citet{Cal:94}.  The correction,
$\Delta\beta$, is obtained for each galaxy with synthetic data equivalent
to Fig.~\ref{fig:dbeta}, but adopting the corresponding best-fit
values of the population and dust parameters, leading to a corrected
(i.e. \textit{Calzetti-equivalent}) $\beta_c$. Fig.~\ref{fig:histobeta} shows the
difference between the distribution of original (orange dashed line) and
corrected UV slopes (red solid line). This comparison shows that
although the changes could be, in principle, rather high, as shown in
Fig.~\ref{fig:dbeta}, the actual properties of the galaxies lead to
changes $\Delta\beta\sim$0.4. The arrows in Fig.~\ref{fig:histobeta}
give the position of the median distribution in \textit{raw} and corrected
$\beta$.

%\betacorrected Rv
\begin{figure}
\centering
\includegraphics[width=85mm]{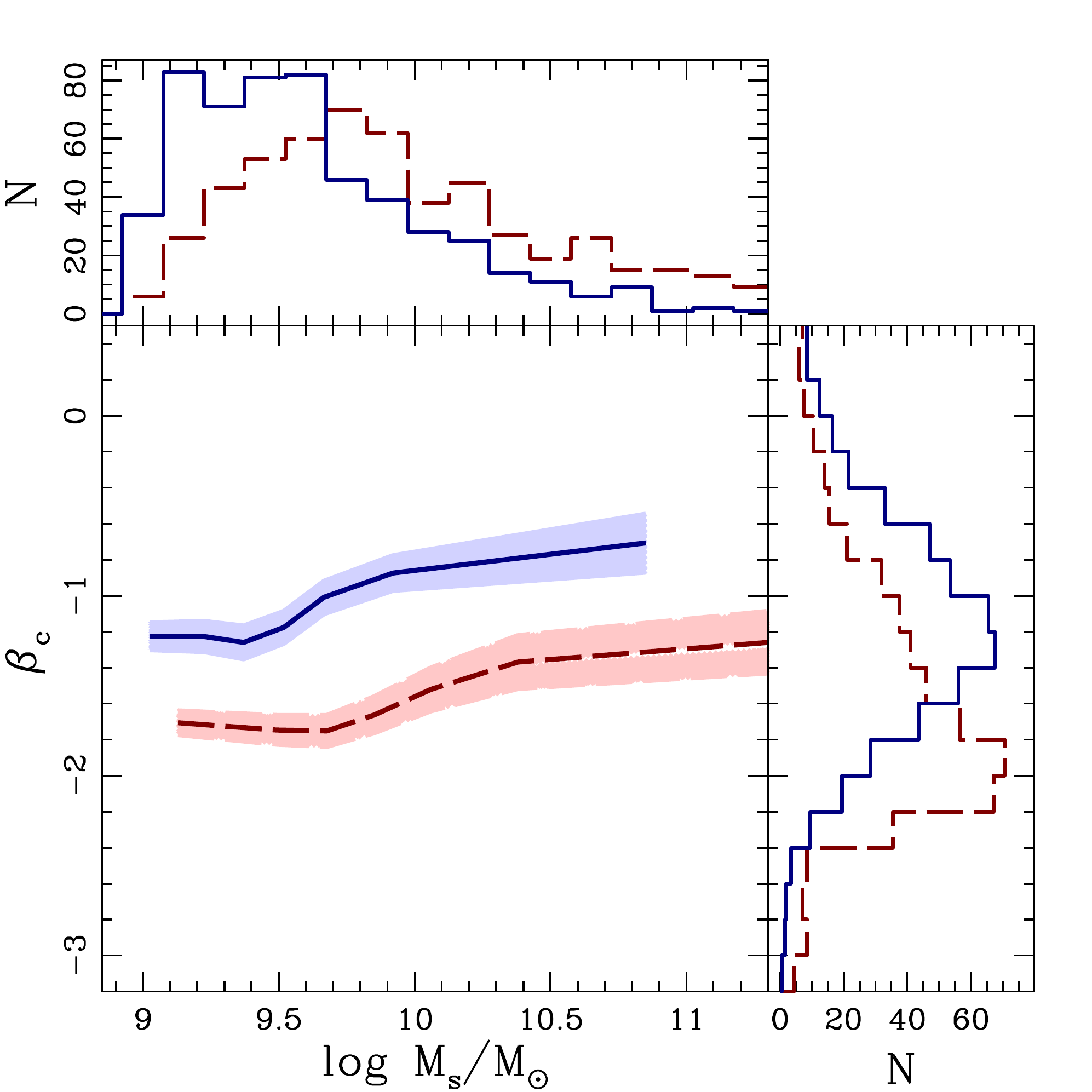}
\caption{The \textit{corrected} UV slope $\beta_{c}$ is plotted 
as a function of the stellar mass.  Histograms for these two
parameters are shown in the top/right panels. In the central plot, the
blue-solid (red-dashed) lines represent objects with ages in the first (third)
tercile of the age distribution, therefore, blue (red) represents
younger (older) populations. The solid lines and shaded regions show
the mean, and its uncertainty, respectively. In the histograms, the
blue (red) lines are also segregated with respect to young (old) ages,
taking the lowest (highest) tercile.}
\label{fig:betacorrmass}
\end{figure}

\begin{figure}
\centering
\includegraphics[width=85mm]{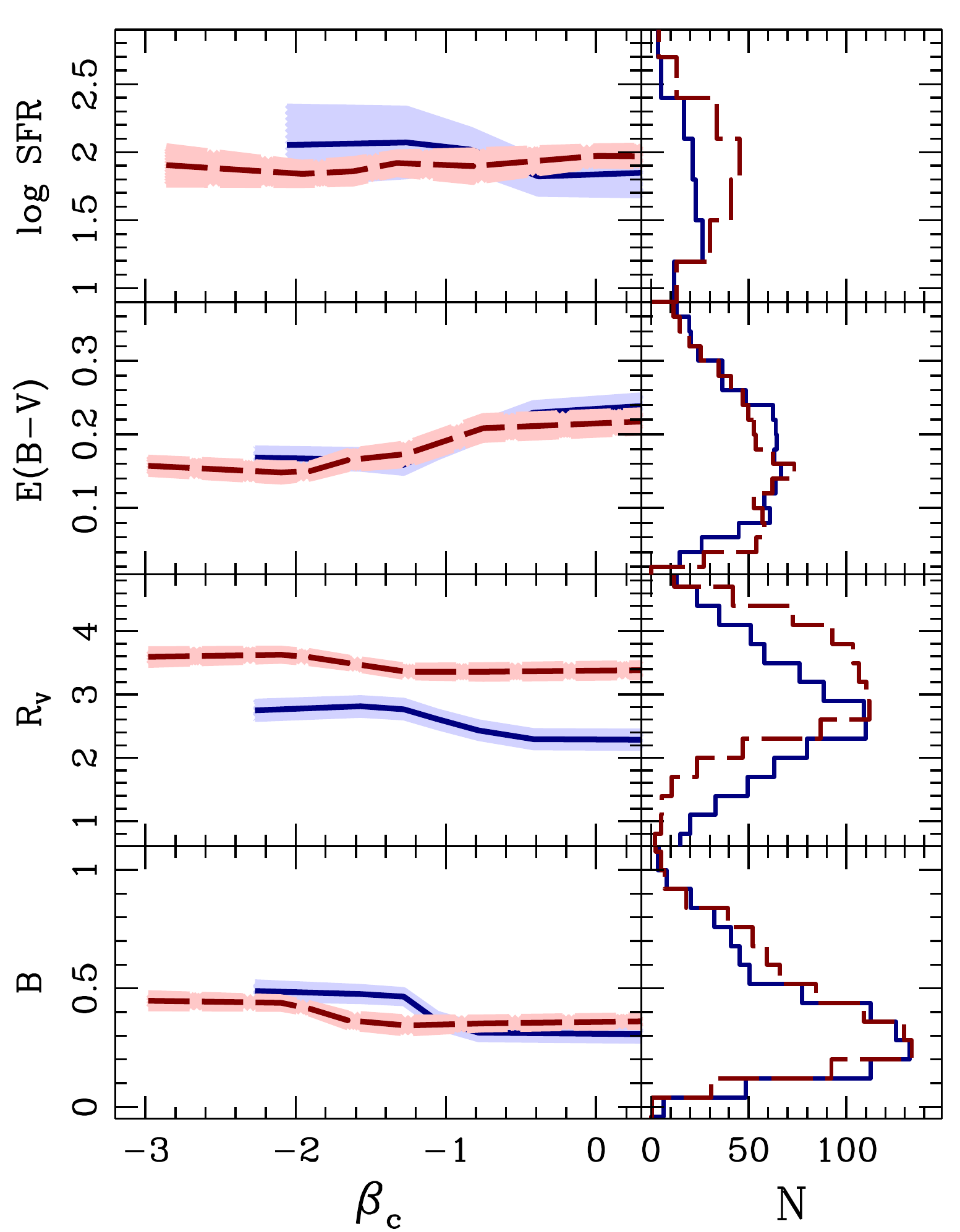}
\caption{The corrected NUV slope, $\beta_{c}$ is shown against
the star formation rate (top) and the dust-related parameters; second
panel from the top, downwards: colour excess, total-to-selective
extinction and NUV bump strength. The line and colour coding follow
the notation in Fig.~\ref{fig:betacorrmass}, split with respect to
age, with red-dashed (blue-solid) representing old (young)
populations.  The panels on the right show the distribution of the
best-fit parameters.}
\label{fig:betacorrdust}
\end{figure}

Figs. \ref{fig:betacorrmass} and \ref{fig:betacorrdust} show the
distribution of $\beta_c$ with respect to several observables of our 
SHARDS sample.  In both figures, the sample is split with respect to age,
showing the median (lines) and median error (shaded regions) of the
youngest (oldest) tercile in stellar age shown in blue-solid (red-dashed).
Fig.~\ref{fig:betacorrmass} shows the relation of UV slope with
stellar mass, with a weak positive correlation in both subsamples.
We include in this figure histograms of the two parameters plotted,
following the same colour coding. The mass histogram (top) shows a
mass-age trend. This is not unexpected, as both stellar metallicity
and age correlate with stellar mass \citep[see, e.g.,][]{GCB:05}. The
$\beta_c$ histogram, to the right of Fig.~\ref{fig:betacorrmass}
reflects the wider scatter in UV slope of the younger subset, but
intriguingly, the youngest population show the flattest UV slopes,
indicative of a higher contribution from dust.

Fig.~\ref{fig:betacorrdust} shows the distribution of $\beta_c$ with
respect to the dust-related parameters. A significant trend is found
between colour excess or bump strength and $\beta_c$, where galaxies
with a shallower UV slope are dustier, and feature weaker bumps. The
trend with colour excess is known for local starbursts \citep{Cal:94}.
When split with respect to age, it is worth mentioning the trend in
R$_V$, where at fixed $\beta_c$, older galaxies have higher R$_V$
(i.e. greyer laws). This trend is also evident in the histograms to
the right of Fig.~\ref{fig:betacorrdust}.

Finally, it is important to note that SSP age, dust content
(quantified through the colour excess) and R$_V$ will induce changes
in the observed UV slope, since changes in these parameters will
affect the shape of the spectral continuum in similar ways. However,
our simulations (\S\ref{Sec:Method}) showed that E($B-V$) and R$_V$
can be retrieved irrespective of this degeneracy.

%\betacorrected Ebv

%%%%%%%%%%%%%%%%%%%%%%%%%%%%%%%%%%%%%%%%%%%%%%%%
\section{Summary}
\label{Sec:Summ}

Taking advantage of the Survey of High-z Absorption Red and Dead
Sources ({\sc SHARDS}), we constrain the dust attenuation law in a
sample of star-forming galaxies over the redshift range
1.5$<$z$<$3. {\sc SHARDS} is an ultra-deep ($<$26.5\,AB, at
4\,$\sigma$) galaxy survey covering 141\,arcmin$^2$ towards the GOODS
North field, that provides optical photo-spectra at resolution
R$\sim$50, via medium band filters (FWHM$\sim$150\AA). Our sample,
selected to have high enough signal in the rest-frame NUV region,
comprises 1,753 galaxies, covering a stellar mass range
9$\simlt\log$\,M/M$_\odot\simlt$11.  We apply a Bayesian method that
explores a wide range of stellar population and dust attenuation
properties. The latter follows the functional form proposed
by \citet{CSB:10}, and is defined by two parameters: the
total-to-selective extinction ratio (R$_V$), and the NUV bump
strength (B), along with a normalization factor, where we use the
colour excess, E($B-V$). A comparison with synthetic models suggests
an accuracy $\Delta$E($B-V$)=0.07; $\Delta$B=0.18; and
$\Delta$R$_V$=0.72 (RMS), without any significant systematic
correlation among the extracted parameters. Furthermore, the
comparison with synthetic models shows that the retrieved values of B,
R$_V$ and E($B-V$) are largely independent of the parameterisation of
the star formation history.

The observational constraints reveal a wide range of values in B and
R$_V$, suggesting complex variations both in the geometric
distribution of dust within galaxies, and/or in the chemical
composition of the dust. A comparison between parameters
(Fig.~\ref{fig:parpar}) shows significant correlations between R$_V$,
B and E($B-V$), towards a steeper attenuation (smaller R$_V$) with
increasing B, and a stronger NUV bump with decreasing colour excess,
quantified by the linear fits shown in eq.~\ref{eq:dust1}
and \ref{eq:dust2}.  This result qualitatively agrees with the study
of \citet{KC:13}, although our results probe a wider range in B and
R$_V$, and we note that our analysis is done on {\sl individual}
galaxies, thanks to the depth of the SHARDS dataset. As expected, we
find a trend between colour excess and stellar mass, and most notably
between SSP-equivalent age and R$_V$, so that the older populations
feature a greyer attenuation (Fig.~\ref{fig:Pars}). We emphasize that
the simulations, including a wide range of star formation histories,
show that the methodology does not introduce any spurious covariance
between these parameters.

We also explored the possibility that the observed wide range of
attenuation parameters could affect the interpretation of the UV
slope. The presence of the NUV bump is especially relevant, since
$\beta$ is measured in a nearby spectral window. Although tests with
simulated data reveal potentially large changes in $\beta$ due to the
NUV bump, we conclude from the observations that realistic corrections
can be of order $\Delta\beta\sim 0.4$.  The UV slope is found to
correlate with the age and the stellar mass of the galaxy
(Fig.~\ref{fig:betacorrmass}). Moreover, at fixed UV slope, older
galaxies feature a higher R$_V$, a consistent trend found throughout
this study.  Although the interpretation of variations in the
attenuation law can be related to either the composition or the
geometric distribution of dust compared to other stars within the
galaxy, this trend with respect to age suggests dust geometry changes
as the main cause, leading us to propose that the relation between
R$_V$ and B shown in the top-left panel of Fig.~\ref{fig:parpar}
(eq.~\ref{eq:dust1}) is caused by variations of clumpiness in the
distribution of dust within star-forming galaxies. We will follow up
this suggestion in a future paper, aimed at breaking the degeneracy
between dust composition and distribution in galaxies, and combining
these results with MIR and FIR data.

\section*{Acknowledgements}
We would like to thank Sandro Bressan for his comments and
suggestions.  M.T. acknowledges support from the Mexican ``Consejo
Nacional de Ciencia y Tecnolog\'\i a'' (CONACyT), with
scholarship \#329741, as well as the Royal Astronomical Society for
travel funding related to this project.  E.M.Q. acknowledges the
support of the European Research Council via the award of a
Consolidator Grant (PI McLure) P.G.P.-G. wishes to acknowledge support
from Spanish Government MINECO Grants AYA2015-63650-P and
AYA2015-70815-ERC.  This work has made use of the Rainbow Cosmological
Surveys Database, which is operated by the Universidad Complutense de
Madrid (UCM), partnered with the University of California
Observatories at Santa Cruz (UCO/Lick, UCSC). Based on observations
made with the GTC, installed at the Spanish Observatorio del Roque de
los Muchachos of the Instituto de Astrof\'\i sica de Canarias, in the
island of La Palma.

%%%%%%%%%%%%%%%%%%%%%%%%%%%%%%%%%%%%%%%%%%%%%%%%
%%%%%%%%%%%%%%%   REFERENCES   %%%%%%%%%%%%%%%%%
%%%%%%%%%%%%%%%%%%%%%%%%%%%%%%%%%%%%%%%%%%%%%%%%

\appendix
\section{Relation between $\delta$,E$_b$ and R$_V$, B}
\label{KCeq}

\begin{figure}
\centering
\includegraphics[width=85mm]{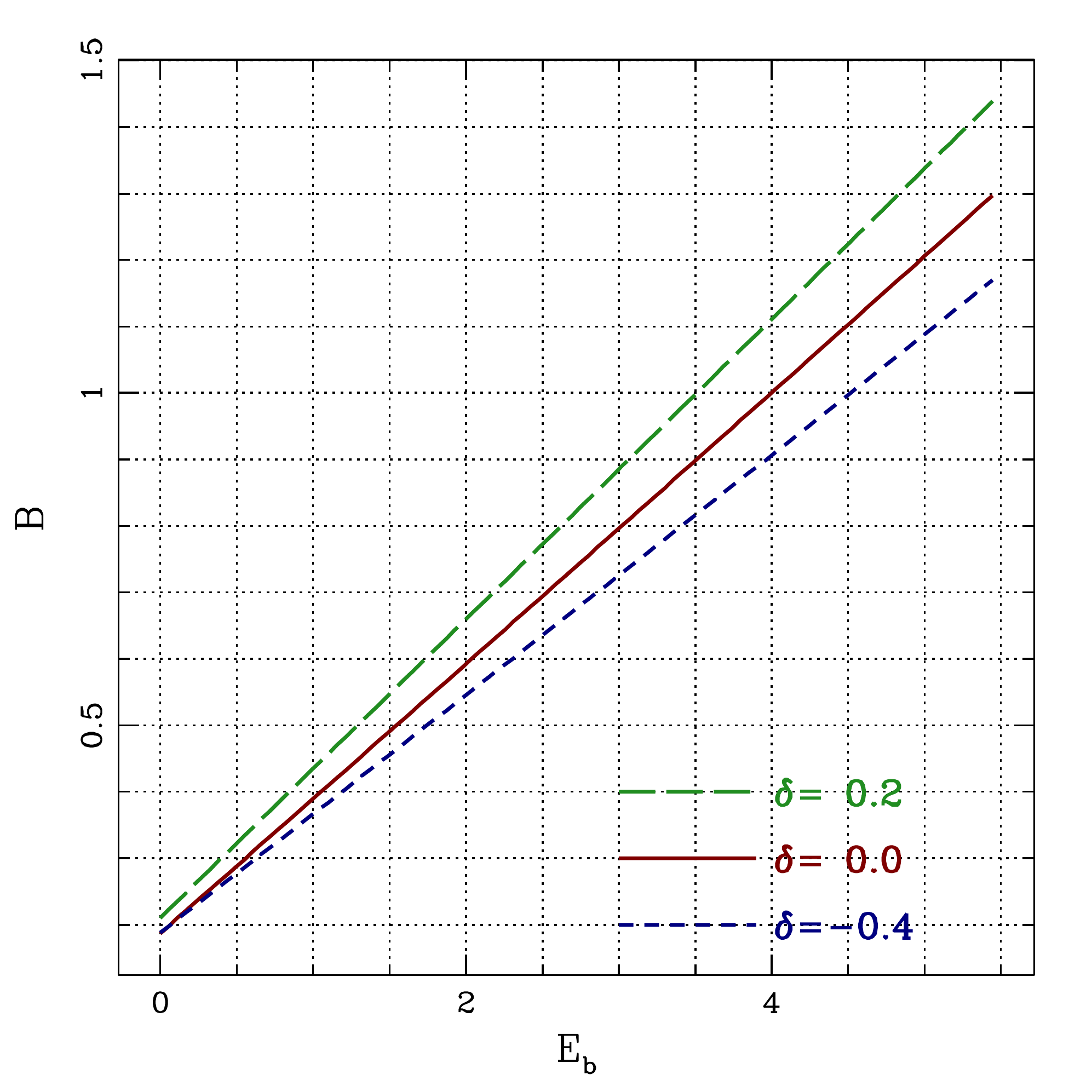}
\caption{Relation between $B$ and E$_b$
for three choices of $\delta$, as labelled.
The linear fits to these trends are
shown in eq.~\ref{eq:ap1}}
\label{fig:BandEb}
\end{figure}

\begin{figure}
\centering
\includegraphics[width=85mm]{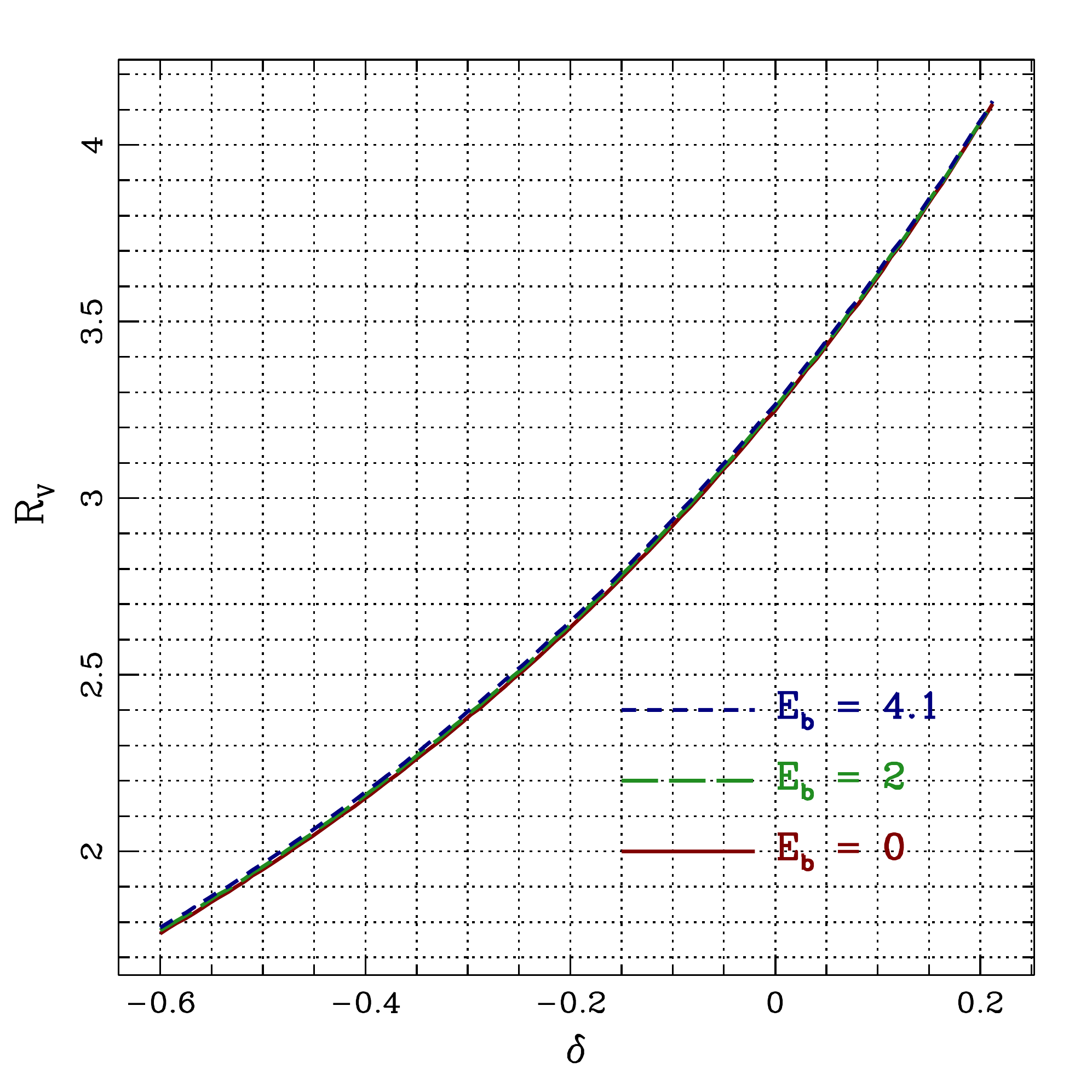}
\caption{Relation between R$_V$ and $\delta$
for three choices of the NUV bump strength parameter E$_b$,
as labelled. The quadratic fit to this trend
(regardless of E$_b$) is shown in eq.~\ref{eq:ap2}}
\label{fig:Rvanddelta}
\end{figure}

We translate the dust parameters $\delta$ and E$_b$ used in the
definition of the attenuation law in \cite{KC:13}, into R$_V$ and $B$
by use of a Python script. The script fits one parameterisation
($\delta$,E$_b$) into the other one (R$_V$,B), within the spectral
window 1,700--6,000\AA.  Fig.~\ref{fig:BandEb}
and \ref{fig:Rvanddelta} show the relation between these two sets of
dust parameters (solid red line).

As expected, the NUV bump strength relation between $E_b$ and B
carries an additional <dependence on $\delta$, since this parameter
``modulates'' the overall wavelength dependence of the attenuation law.
Fig.~\ref{fig:BandEb} shows the results for three choices of
$\delta$, as labelled. A linear fit to these lines give:
\begin{equation}
\label{eq:ap1}
\left.
\begin{aligned}
B &= 0.2255 E_b + 0.2091 \qquad \delta=+0.2\\
B &= 0.2038 E_b + 0.1856 \qquad \delta=0\\
B &= 0.1802 E_b + 0.1861 \qquad \delta=-0.4\\
\end{aligned}
\right\}
\end{equation}

The relationship between $\delta$ and R$_V$, shown in
Fig.~\ref{fig:Rvanddelta} is rather unsensitive to the
choice of the bump strength, although it requires a
higher order polynomial. We find a quadratic fit:
\begin{equation}
\label{eq:ap2}
R_V = 1.747\delta^2 + 3.503\delta + 3.271
\end{equation}
gives a good representation, independently of the value of E$_b$.
Note that in the parameterisation of \citet{CSB:10}, the
``Milky Way equivalent'', $R_V=3.1$, $B=1$ is mapped
into $\delta\simeq-0.05$, $E_b\simeq 4.0$.

%B and Eb

%Rv and \delta

\label{lastpage}


\begin{thebibliography}{}

\bibitem[\protect\citeauthoryear{Alexander et al.}{2003}]{cdfn}
Alexander D.~M., et al., 2003, AJ, 126, 539 

\bibitem[\protect\citeauthoryear{Barro et al.}{2011}]{Barro:11}
Barro G., et al., 2011, ApJS, 193, 13 

\bibitem[\protect\citeauthoryear{Battisti et al.}{2016}]{BCC:16}
Battisti A.~J., Calzetti D., Chary R.-R., 2016, ApJ, 818, 13 

\bibitem[\protect\citeauthoryear{Bekki et al.}{2015}]{BHT:15}
Bekki K., Hirashita H., Tsujimoto T., 2015, ApJ, 810, 39 

\bibitem[\protect\citeauthoryear{Bradley et al.}{2005}]{Bradley:05}
Bradley J., et al., 2005, Sci, 307, 244 

\bibitem[\protect\citeauthoryear{Bruzual \& Charlot}{2003}]{BC03} 
Bruzual, G., Charlot., S., 2003, MNRAS, 344, 1000

\bibitem[\protect\citeauthoryear{Buat et al.}{2011}]{BGH:11}
Buat V., et al., 2011, A\&A, 533, A93 

\bibitem[\protect\citeauthoryear{Buat et al.}{2012}]{BNB:12}
Buat V., et al., 2012, A\&A, 545, A141 

\bibitem[\protect\citeauthoryear{Burgarella et al.}{2005}]{BBI:05}
Burgarella D., Buat V., Iglesias-P{\'a}ramo J., 2005, MNRAS, 360, 1413 

\bibitem[\protect\citeauthoryear{Calzetti et al.}{1994}]{Cal:94}
Calzetti D., Kinney A.~L., Storchi-Bergmann T., 1994, ApJ, 429, 582 

\bibitem[\protect\citeauthoryear{Calzetti \& Heckman}{1999}]{Cal:99}
Calzetti D., Heckman T.~M., 1999, ApJ, 519, 27 

\bibitem[\protect\citeauthoryear{Calzetti et al.}{2000}]{Cal:00}
Calzetti D., Armus L., Bohlin R.~C., Kinney A.~L., Koornneef J.,
Storchi-Bergmann T., 2000, ApJ, 533, 682 

\bibitem[\protect\citeauthoryear{Cardelli et al.}{1989}]{CCM:89} 
Cardelli J.~A., Clayton G.~C., Mathis J.~S., 1989, ApJ, 345, 245

\bibitem[\protect\citeauthoryear{Cava et al.}{2015}]{Cava:15}
Cava A., et al., 2015, ApJ, 812, 155 

\bibitem[\protect\citeauthoryear{Chabrier}{2003}]{Chabrier:03}
Chabrier G., 2003, PASP, 115, 763 

\bibitem[\protect\citeauthoryear{Charlot \& Fall}{2000}]{CF:00}
Charlot, S., Fall., S.~M., 2000, ApJ, 539, 718
 
\bibitem[\protect\citeauthoryear{Chevallard et al.}{2013}]{Chevallard:13}
Chevallard, J., Charlot, S., Wandelt, B., Wild, V., 2013, MNRAS, 432, 2061

\bibitem[\protect\citeauthoryear{Conroy et al.}{2010}]{CSB:10}
Conroy, C., Schiminovich, D., Blanton, M.~R., 2010, ApJ, 718, 184

\bibitem[\protect\citeauthoryear{Conroy}{2010}]{C:10}
Conroy C., 2010, MNRAS, 404, 247 

\bibitem[\protect\citeauthoryear{Dom{\'{\i}}nguez S{\'a}nchez et al.}{2016}]{HDS:16}
Dom{\'{\i}}nguez S{\'a}nchez H., et al., 2016, MNRAS, 457, 3743 

\bibitem[\protect\citeauthoryear{Donley et al.}{2012}]{Donley:12}
Donley, J.~L., et al. 2012, ApJ, 748, 142

\bibitem[\protect\citeauthoryear{Draine}{1989}]{Draine:89}
Draine B., 1989, IAUS, 135, 313 

\bibitem[\protect\citeauthoryear{Draine}{2003}]{Draine:03}
Draine B.~T., 2003, ARA\&A, 41, 241 

\bibitem[\protect\citeauthoryear{Fazio et al.}{2004}]{IRAC}
Fazio G.~G., et al., 2004, ApJS, 154, 10 

\bibitem[\protect\citeauthoryear{Ferreras et al.}{2012}]{FW4871}
Ferreras I., et al., 2012, AJ, 144, 47 

\bibitem[\protect\citeauthoryear{Ferreras et al.}{2014}]{Ferreras:14}
Ferreras I., et al., 2014, MNRAS, 444, 906 

\bibitem[\protect\citeauthoryear{Fitzpatrick}{1999}]{F:99}
Fitzpatrick E.~L., 1999, PASP, 111, 63

\bibitem[\protect\citeauthoryear{Gallazzi et al.}{2005}]{GCB:05}
Gallazzi A., Charlot S., Brinchmann J., White S.~D.~M., Tremonti C.~A., 2005, MNRAS, 362, 41 

\bibitem[\protect\citeauthoryear{Galliano et al.}{2017}]{Galliano:17}
Galliano, F., Galametz, M., Jones, A.~P., 2017, ARA\&A, in press, arXiv:1711.07434

\bibitem[\protect\citeauthoryear{Giavalisco et al.}{2004}]{ACS}
Giavalisco, M., et al., 2004, ApJ, 600, L93

\bibitem[\protect\citeauthoryear{Gordon, Smith, \& Clayton}{1999}]{GSC:99}
Gordon K.~D., Smith T.~L., Clayton G.~C., 1999, ASPC, 193, 517 

\bibitem[\protect\citeauthoryear{Hagen et al.}{2017}]{HSH:16}
Hagen L.~M.~Z., Siegel M.~H., Hoversten E.~A., Gronwall C., Immler S.,
Hagen A., 2017, MNRAS, 466, 4540

\bibitem[\protect\citeauthoryear{Hutton et al.}{2014}]{HF:14}
Hutton S., Ferreras I., Wu K., Kuin P., Breeveld A., Yershov V.,
Cropper M., Page M., 2014, MNRAS, 440, 150

\bibitem[\protect\citeauthoryear{Hutton et al.}{2015}]{HF:15}
Hutton S., Ferreras I., Yershov V., 2015, MNRAS, 452, 1412 

\bibitem[\protect\citeauthoryear{Johnson et al.}{2007}]{JSS:07}
Johnson B.~D., et al., 2007, ApJS, 173, 392 

\bibitem[\protect\citeauthoryear{Kajisawa et al.}{2011}]{Kasijawa:11}
Kajisawa M., et al., 2011, PASJ, 63, 379 

\bibitem[\protect\citeauthoryear{Koekemoer et al.}{2011}]{CANDELS}
Koekemoer A.~M., et al., 2011, ApJS, 197, 36 

\bibitem[\protect\citeauthoryear{Kriek \& Conroy}{2013}]{KC:13}
Kriek M. and Conroy C. 2013 ApJ, 775, L16

\bibitem[\protect\citeauthoryear{McCracken et al.}{2010}]{CFHT}
McCracken, H.J., 2010, ApJ, 708, 202

\bibitem[\protect\citeauthoryear{Murata et al.}{2014}]{Mu:14}
Murata et al., 2014, A\&A, 566, 136

\bibitem[\protect\citeauthoryear{Leitherer et al.}{1999}]{SB99}
Leitherer C., et al., 1999, ApJS, 123, 3 

\bibitem[\protect\citeauthoryear{Noll et al.}{2009}]{NPC:09}
Noll S., et al., 2009, A\&A, 499, 69 

\bibitem[\protect\citeauthoryear{Panuzzo et al.}{2007}]{PGB:07}
Panuzzo P., Granato G.~L., Buat V., Inoue A.~K., Silva L.,
Iglesias-P{\'a}ramo J., Bressan A., 2007, MNRAS, 375, 640 

\bibitem[\protect\citeauthoryear{Papovich et al.}{2011}]{Papovich:11}
Papovich C., Finkelstein S.~L., Ferguson H.~C., Lotz J.~M., Giavalisco
M., 2011, MNRAS, 412, 1123

\bibitem[\protect\citeauthoryear{Pei}{1992}]{Pei:92}
Pei Y.~C., 1992, ApJ, 395, 130 

\bibitem[\protect\citeauthoryear{P{\'e}rez-Gonz{\'a}lez et al.}{2013}]{SHARDS}
P{\'e}rez-Gonz{\'a}lez P.~G., et al., 2013, ApJ, 762, 46 

\bibitem[\protect\citeauthoryear{P{\'e}rez-Gonz{\'a}lez et al.}{2008}]{PG:08}
P{\'e}rez-Gonz{\'a}lez P.~G., et al., 2008, ApJ, 675, 234-261 

\bibitem[\protect\citeauthoryear{Reddy et al.}{2012}]{Reddy:12}
Reddy N.~A., Pettini M., Steidel C.~C., Shapley A.~E., Erb D.~K., Law
D.~R., 2012, ApJ, 754, 25

\bibitem[\protect\citeauthoryear{Reddy et al.}{2015}]{RKS:15}
Reddy N.~A., et al., 2015, ApJ, 806, 259 

\bibitem[\protect\citeauthoryear{Rujopakarn et al.}{2013}]{Rujopakarn:13}
Rujopakarn W., Rieke G.~H., Weiner B.~J., P{\'e}rez-Gonz{\'a}lez P.,
Rex M., Walth G.~L., Kartaltepe J.~S., 2013, ApJ, 767, 73

\bibitem[\protect\citeauthoryear{Salmon et al.}{2016}]{SPL:16}
Salmon B., et al., 2016, ApJ, 827, 20 

\bibitem[\protect\citeauthoryear{Scoville et al.}{2015}]{SFC:15}
Scoville N., Faisst A., Capak P., Kakazu Y., Li G., Steinhardt C., 2015, ApJ, 800, 108 

\bibitem[\protect\citeauthoryear{Seon \& Draine}{2016}]{SD:16}
Seon K.-I., Draine B.~T., 2016, ApJ, 833, 201 

\bibitem[\protect\citeauthoryear{Shivaei et al.}{2017}]{SKR:16}
Shivaei I., et al. ApJ 820 L23

\bibitem[\protect\citeauthoryear{Speagle et al.}{2014}]{Speagle:14}
Speagle J.~S., Steinhardt C.~L., Capak P.~L., Silverman J.~D., 2014,
ApJS, 214, 15

\bibitem[\protect\citeauthoryear{Stecher}{1969}]{STP:69}
Stecher T.~P., 1969, ApJ, 157, L125 

\bibitem[\protect\citeauthoryear{Trouille et al.}{2008}]{Trouille:08}
Trouille L., Barger A.~J., Cowie L.~L., Yang Y., Mushotzky R.~F.,
2008, ApJS, 179, 1

\bibitem[\protect\citeauthoryear{Wu et al.}{2006}]{Wu:06}
Wu Y., Charmandaris V., Hao L., Brandl B.~R., Bernard-Salas J., Spoon
H.~W.~W., Houck J.~R., 2006, ApJ, 639, 157

\bibitem[\protect\citeauthoryear{Wild et al.}{2011}]{WCB:11}
Wild V., Charlot S., Brinchmann J., Heckman T., Vince O., Pacifici C.,
Chevallard J., 2011, MNRAS, 417, 1760

\bibitem[\protect\citeauthoryear{Witt \& Gordon}{2000}]{WG:00}
Witt, A.~N., Gordon, K.~D., 2000, ApJ, 528, 799

\bibitem[\protect\citeauthoryear{Zahid et al.}{2013}]{ZYKK:13}
Zahid H.~J., Yates R.~M., Kewley L.~J., Kudritzki R.~P., 2013, ApJ,
763, 92

\bibitem[\protect\citeauthoryear{Zeimann et al.}{2015}]{ZG:15}
Zeimann G.~R., et al., 2015, ApJ, 814, 162 


%%%%%%%%%%%%%%%%%%


\end{thebibliography}
\end{document}